\documentclass[useAMS,usenatbib,usegraphicx]{mn2e}
\usepackage{supertabular,lscape,rotating}

\newcommand{\apj}{ApJ}
\newcommand{\apjs}{ApJS}
\newcommand{\aap}{A\&A}
\newcommand{\aaps}{A\&AS}

\newcommand{\degs}{$^{\circ}$}
\newcommand{\hii}{H\,{\sc ii}}
\newcommand{\hi}{H\,{\sc i}}

\title[Resolving the Kinematic Distance Ambiguity of Southern Massive Young Stellar Object Candidates]{Resolving the Kinematic Distance Ambiguity of Southern Massive Young Stellar Object Candidates}
\author[A. L. Busfield et al.]{A. L. Busfield,$^{1}$\thanks{E-mail: alb,mgh,sll,roud@ast.leeds.ac.uk} C. R. Purcell,$^{2}$\thanks{E-mail: crp@unsw.edu.au} M. G. Hoare,$^{1}$\footnotemark[1] S. L. Lumsden,$^{1}$\footnotemark[1] \newauthor T. J. T. Moore$^{3}$\thanks{E-mail: tjtm@astro.livjm.ac.uk} and R. D. Oudmaijer$^{1}$\footnotemark[1] \\
$^{1}$School of Physics and Astronomy, University of Leeds, Leeds, LS2 9JT, UK \\
$^{2}$School of Physics, University of New South Wales, Sydney, Australia 2052 \\
$^{3}$Astrophysics Research Institute, Liverpool John Moores University, Twelve Quays House, Egerton Wharf, Birkenhead CH41 1LD, UK}

\begin{document}

\date{Accepted Received }

\pagerange{\pageref{firstpage}--\pageref{lastpage}} \pubyear{2003}

\maketitle

\label{firstpage}

\begin{abstract}

We investigate the use of \hi\ data to resolve the near/far ambiguity
in kinematic distances of massive young stellar object (MYSO)
candidates. Kinematic distances were obtained from $^{13}$CO 1-0 (and
N$_{2}$H$^{+}$) spectral line observations with the Mopra Telescope
towards 94 candidates selected from the Red \textit{MSX} Source (RMS)
survey in the fourth Galactic quadrant ($ 282^{\circ} < l <
350^{\circ}$). \hi\ data from the Southern Galactic Plane Survey
(SPGS) was used in conjunction with the \hi\ self-absorption technique
to determine the near or far distance. We resolved the kinematic
distance ambiguity to 70\% of the sources. We can also simultaneously
solve for any multiple line-of-sight component sources. We discuss the
advantages and disadvantages of this technique in comparison with
other methods, and also perform confidence checks on the reliability
of using the \hi\ self-absorption technique.

We examined the projected location of these sources in both the
Galactic plane and longitude-velocity diagrams to ascertain any
recognisable spiral arm pattern. Although no obvious spiral pattern
was found when compared to that proposed by Cordes \& Lazio, far
distance sources tended to lie on or near spiral arm loci. Near
distance sources, however, had peculiar velocity uncertainties
consistent with the separation between the spiral arms themselves. The
longitude-velocity plot shows a more consistent picture, with tangent
points of the spiral arm loci easily seen.

We conclude that using the \hi\ self-absorption technique to determine
kinematic distance ambiguities is a quick and reliable method in most
cases, with an 80\% success rate in determining the correct
designation of whether an object is at the near or far distance.

\end{abstract}


\section{Introduction}

Stars with mass $>$~8~M$_{\sun}$ (L $> 10^4$~L$_{\sun}$), play
important roles in many areas of astrophysics. They are fundamental to
the evolution of galaxies, because they produce the heavy elements,
inject large amounts of UV radiation, wind energy and strong shocks
into the interstellar medium throughout their short lifetimes, and
possibly regulate the star formation rate. Even though much has been
learnt over the past two decades about the evolution of massive stars,
their birth still remains mysterious.

Many aspects of the study of massive star formation suffer from the
lack of a large unbiased sample. In particular the observationally
known sample of massive young stellar objects (MYSOs) only consists of
thirty or so well catalogued objects \citep{24}.  Formed just after
the hot core phase, MYSOs are mid-IR bright objects where fusion has
probably begun within the core, however, it has not yet started to
ionise the surroundings to form an ultra-compact \hii\ region.  Many
of the known MYSOs are nearby and may not represent this class of
objects as a whole. Hence the need for finding a larger sample of
MYSOs within our Galaxy, using an unbiased selection technique.

We have developed colour-selection criteria that deliver about 2000
candidate MYSOs from the \textit{MSX} point source catalogue with the
addition of 2MASS data \citep{1}. There are many other types of source
which potentially have the same very red colours as MYSOs, especially
compact \hii\ regions, but also some planetary nebulae, post-AGB
stars, dusty red supergiants and low mass YSOs. A large follow-up
programme, the Red \textit{MSX} Source or RMS Survey \citep{19} is
underway to identify and eliminate these other sources from the
sample.  A key part of the follow-up is the determination of distances
to these sources in order to obtain bolometric luminosities. In fact,
some of them might be low luminosity YSOs and therefore should be
removed from the sample.

The determination of kinematic distances only requires the measurement
of the radial velocity of the source together with the application of
a specific Galactic rotation curve (e.g. Brand \& Blitz
1993). Although the determination of kinematic distances in the outer
Galaxy is relatively trivial, it becomes much more difficult towards
the inner Galaxy due to the distance ambiguity problem. In the inner
Galaxy a single radial velocity value relates to two distances (a near
and far distance) either side of the tangent point, along any one line
of sight within the solar circle. Only sources that lie at the tangent
point have unambiguous distances.




Many molecular line studies have been used to search for massive young
stars, dense molecular cores or \hii\ regions, and determine their
distances \citep{10,11,7}. However, the majority of these studies have
side-stepped the distance ambiguity problem altogether by observing
sources outside the solar circle (e.g. Wouterloot \& Brand 1989) hence
ignoring the inner Galaxy where the bulk of massive star formation
occurs. A number of studies where inner Galaxy objects have been
observed have either been left with the unsolved ambiguity problem and
no kinematic distances calculated (e.g. Sridharan \textit{et al.}
2002), or have assumed all objects with an ambiguity to be at a near
or a far distance, resulting in bolometric luminosities which can be
orders of magnitude different (e.g. Wu, Wu \& Wang 2001).

Several studies have tried to solve the kinematic distance ambiguity
to a small number of sources using various
techniques. \citet{10,13,12} have combined measurements of the
H110$\alpha$ radio recombination line (RRL) and the H$_2$CO
(formaldehyde) absorption line to distinguish between near and far
distances. To determine kinematic distances using this method, it is
assumed that all H$_2$CO absorption lines are produced by absorption
of radio continuum from a single background \hii\ region. Based on
these assumptions, if the H$_2$CO absorption occurs at higher
velocities than the RRL, then the \hii\ region has to be located at
the far distance. If no molecular gas is detected at velocities
higher than the RRL, then the near kinematic distance is implied for
the \hii\ region. Although this method is useful, it can only be used
for radio bright sources, whereas the MYSOs are known to be radio
quiet.

A more statistical approach was made by \citet{25} using the
luminosity-physical diameter correlation of \hii\ regions. The
relationship is based on the argument that the luminosity is
proportional to the product of the emission measure (which is
proportional to the linear diameter) with the square of the angular
size. A highly significant relationship was found, and resolved the
distance ambiguity to 256 sources. However, this method is again only
able to be used for well behaved \hii\ regions, and is only successful
in solving 60\%\ of re-tested cases where the distance ambiguity had
already been resolved.

\citet{5} and \citet{4}(hereafter LBB) identified several \hi\
self-absorption (SA) features in 21 cm \hi\ maps made of the Galactic
plane and noted that many of these also correlate with CO emission
features. LBB hypothesised that \hi\ SA may provide a robust method
for determining the distance to molecular clouds. If the cold \hi\ in
a molecular cloud absorbs warmer \hi\ background line emission, then
clouds at the near distance should show \hi\ SA, whereas clouds at the
far distance will not since there there is no background to
absorb. \citet{3} solved the distance ambiguity to the molecular cloud
GRSMC 45.6$+$0.3 using \hi\ self-absorption (SA). This approach
requires all molecular clouds have to contain sufficient cold atomic
hydrogen in order for \hi\ SA to occur. Although this still needs to
be confirmed observationally, \citet{28} have shown that all their
model molecular clouds contain enough opacity in cold \hi\ to exhibit
self-absorption against strong 21 cm backgrounds.

\begin{figure}
\vspace{0.5cm}
\hspace{-0.5cm}
\includegraphics[width=85mm]{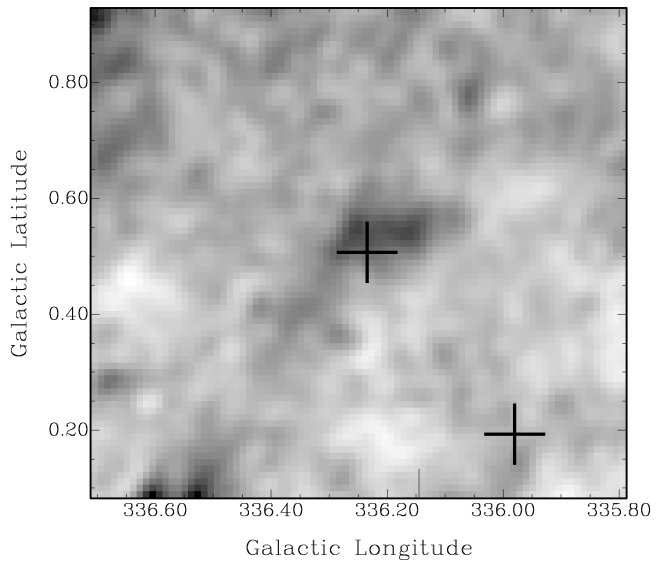}
 \caption{An example \hi\ cube slice (greyscale) centered at the $v = -29.7~km~s^{-1}$ velocity channel overlaid with the central position of one of our \textit{MSX} targets G336.2344+00.5067, showing a slightly extended darker (absorption) region around the object. The other black cross indicates another RMS potential MYSO candidate within the field at a different velocity.}
  \label{movie}
\end{figure}

Up to now, \hi\ data with spatial resolution comparable to CO surveys
and covering the whole Galaxy were not available, with only small
samples of the Galactic plane surveyed such as the Boston
University-Arecibo Observatory (BUAO) Galactic \hi\ Survey
\citep{22}. These data have since been successfully used with the \hi\
SA method on a small selection of objects by \citet{14}. However, with
the near completion of the International Galactic Plane Survey (IGPS),
it is now possible to use this method to study the entire
Galaxy. Further details about this survey will be given in the next
section.

Many of our candidate MYSOs have now been observed at $^{13}$CO 1-0 to
obtain kinematic distances and hence bolometric luminosities. Unlike
$^{12}$CO 1-0, $^{13}$CO 1-0 has narrower line widths allowing much
better separation of the velocity components. Furthermore it also
avoids some of the difficulties of self-absorbed line profiles often
found with $^{12}$CO data especially towards the inner Galaxy reducing
the assignment of an incorrect spectral line velocity.

The source list was compiled primarily from the second release of the
\textit{MSX} point source catalogue, using colour-selection criteria
for MYSOs from \citet{1} combined with visual inspection of all the
\textit{MSX} images for likely targets, delivering 2000 objects. This
paper will concentrate on a sample of these objects with a distance
ambiguity in the 4th Galactic quadrant and it will discuss whether the
\hi\ SA technique is a valid method for determining their kinematic
distance.

\section{Observations and Data Reduction}

The $^{13}$CO 1-0 observations were carried out in the periods of
August 2002 and July 2003, using the Mopra 22-m millimetre-wave
telescope of the ATNF. The assumed rest frequency of the $^{13}$CO 1-0
transition was 110.201353 GHz. The half-power beamwidth of the Mopra
antenna at $^{13}$CO 1-0 is $\sim$35\arcsec. The back-end was a 64 MHz
wide 1024-channel autocorrelator, which yields a channel resolution at
110 GHz of 0.17 km s$^{-1}$.

The observations were performed in a position-switching mode with
reference positions offset 1 degree away from the object, in a
direction away from the plane of the Galaxy. Total on-source time was
usually 10 minutes, using 10$\times$1 minute integrations; less if the
lines were strong, giving a rms noise of $\sim$0.1 K. The reference
position was chosen to try and minimise any off-beam emission from
appearing at the on-source velocity within the crowded Galactic
plane. The antenna pointing was checked and corrected every 2-3 hours
by making observations of strong 86 GHz SiO masers; the pointing
accuracy when this procedure is followed is $\sim$10 arcsec.

A cryogenically cooled low-noise SIS mixer was used in the
receiver. The system temperature varied between 240 and 400 K,
depending on weather conditions and the elevation of the
telescope. Although we were only primarily interested in obtaining a
kinematic velocity, and not accurate fluxes, an ambient temperature
load (assumed to have a temperature of 290 K) was regularly placed in
front of the receiver to enable calibration of the observed flux
density for the effects of atmospheric attenuation. The estimated
uncertainty of the absolute flux density scale is 20 per cent.

The $^{13}$CO 1-0 spectral line data were reduced in the standard way
with \textsc{dfm} (Data From Mopra)\footnote[1] {A tcl/tk graphical
interface written for \textsc{spc} (Spectral Line Reduction Package),
by C. Purcell.}.  The separate integrations were averaged, and
baseline subtracted. To improve the signal-to-noise ratio of weak
sources, the spectra were then Hanning smoothed. Gaussians were fitted
to all line components, and any significant off-beam sources that do
not affect the line profile zeroed. The uncertainty in fitting the
velocity profiles was typically ~$\pm$ 0.1 km$^{-1}$

Where there are multiple $^{13}$CO emission components along a single
line of sight, we have identified the likely MYSO component to be the
strongest and/or widest component within the spectra. However, this
may not always be correct. If there is no broad component more than
three times the strength of the others, the line with a width greater
than 5 km s$^{-1}$ was used. It should be noted that not all of these
objects will be star-forming regions and the RMS survey is currently
underway, identifying all these regions.

The IGPS is a collaboration of the Southern (SGPS) \citep{2}, VLA
(VGPS) \citep{20}, and Canadian (CGPS) \citep{21} Galactic Plane
Surveys.  The \hi\ data cubes were obtained from the Southern Galactic
Plane Survey of \citet{2}. The SGPS consisted of 2212 mosaic pointings
over a total area of 210 square degrees, with an angular resolution of
2\arcmin\ and a rms line noise level of $\sim$ 2.3K. The
continuum-subtracted cubes were viewed using the \textsc{karma}
\footnote[2] {\textsc{karma} is a toolkit for interprocess
communications, authentication, encryption, graphics display, user
interface and manipulating the \small{KARMA} network data structure,
written by R. Gooch.}  general-purpose image display software. The
spectrum from the relevant object pixel position were extracted from
the \hi\ data cubes (Figure~\ref{movie}), with a pixel size of
$\sim$0.02\degs. There were no significant spectral deviations from
the central object pixel and its surroundings on any of the \hi\
spectra.

\begin{figure*}
\begin{minipage}{175mm}
\hspace{-0.5cm}
\includegraphics[width=180mm,height=80mm]{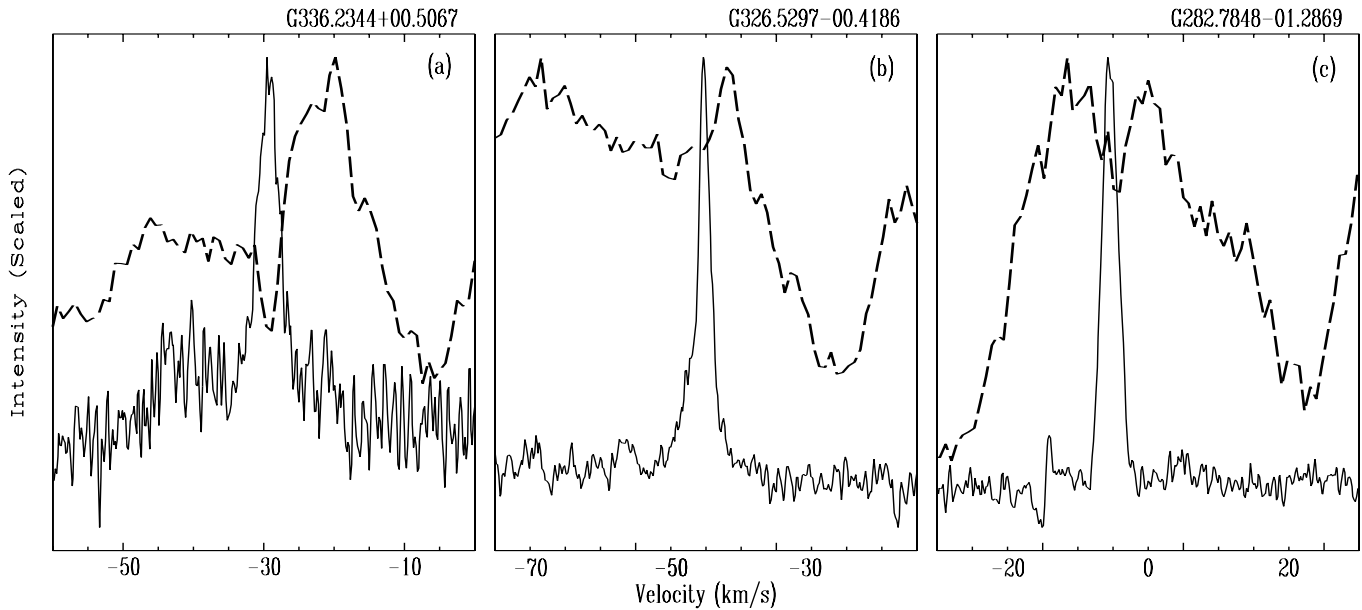}
 \caption{Representative spectra showing the three possible solutions obtained using the \hi\ self-absorption method. The solid line represents the $^{13}$CO spectra overlaid with the \hi\ 21cm data (bold dashed line), with the \hi\ data scaled to the peak of $^{13}$CO. Plot (a) shows a clear near distance solution with the $^{13}$CO emission line coinciding with a \hi\ absorption. Plot (b) represents a far distance object with no \hi\ SA located at the $^{13}$CO peak velocity, whereas plot (c) shows an example of an ambiguity, as overall the $^{13}$CO peak lies in a \hi\ trough, but within the trough is a spike, hence not providing any clear cut answer to the kinematic distance ambiguity.}
  \label{examples}
\end{minipage}
\end{figure*}

For five of the RMS survey objects we use the N$_{2}$H$^{+}$
transition line instead of the $^{13}$CO 1-0 line to determine the
kinematic distance, using data obtained in a similar way at Mopra by
Purcell et al (in prep.). These RMS survey objects are sites of
methanol maser emission, thought to be tracers of massive star
formation. However, due to the spread in maser velocities (up to
$\sim$10 km s$^{-1}$) we are using the thermal emission line to obtain
the object velocity. Due to hyperfine splitting of N$_{2}$H$^{+}$
line, the position of the central peak was used to determine the
velocity.

\section{Results and Discussion}

\begin{figure}
\vspace{0.5cm}
\hspace{-0.7cm}
\includegraphics[width=90mm,height=139mm]{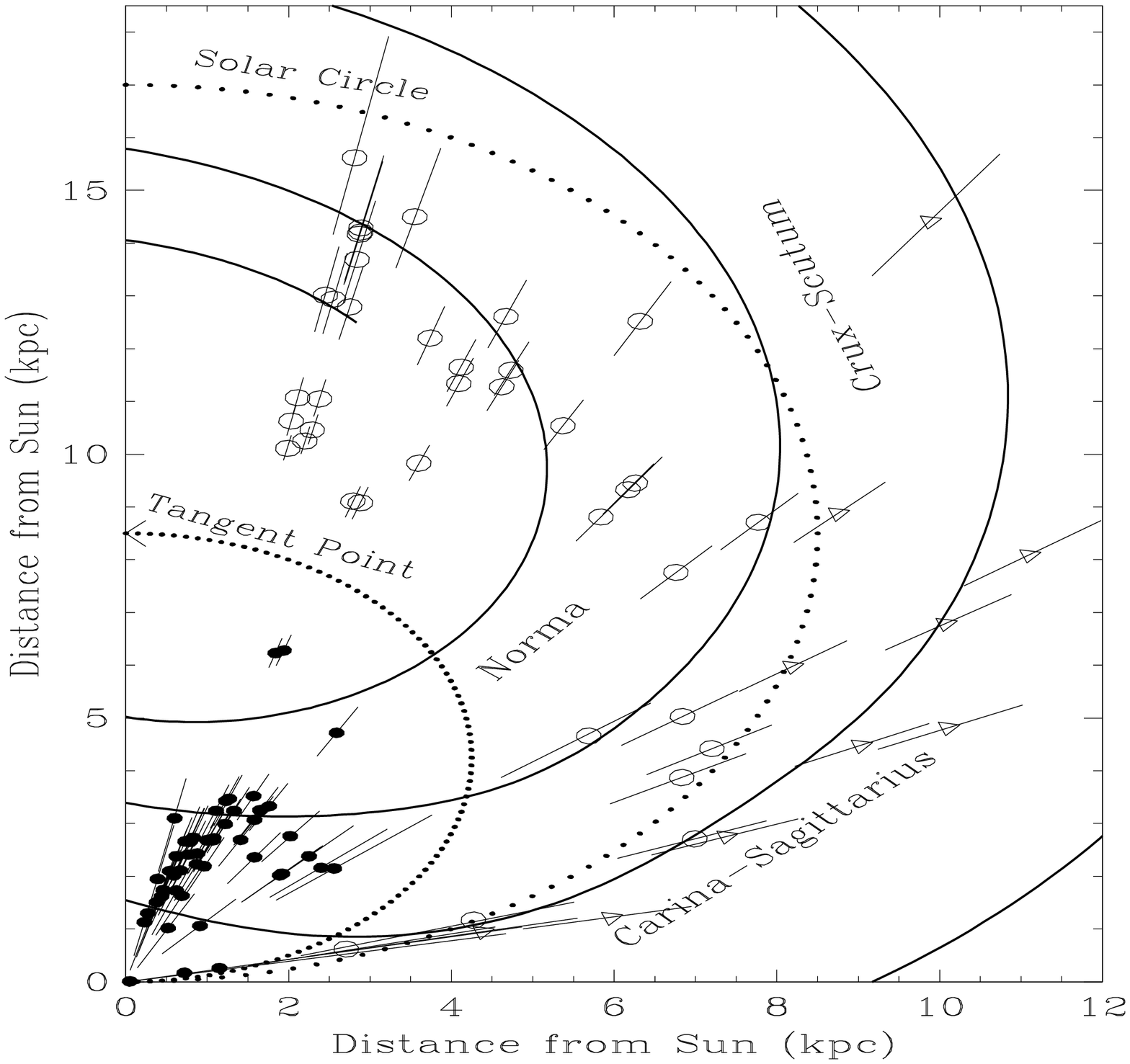}
 \caption{Galactic plot showing the distance ambiguity problem after comparison to the \hi\ data. Filled circles in the inner (tangent) circle represent near distances, whereas the open circles represent far distances within in the outer (solar) circle. Open triangles represent the 10 sources which were studied with no distance ambiguity, and a typical peculiar velocity error of $\pm$ 10 km s$^{-1}$ is indicated on each point. The updated spiral arm model of \citet{17} by \citet{16} is indicated by solid lines. }
  \label{galaxy}
\end{figure}

Figure~\ref{examples} shows detailed examples of the various types of
$^{13}$CO and overlaid \hi\ spectral profiles that were obtained.
Figure~\ref{examples} (a) shows a clear near distance solution with
the $^{13}$CO emission line coinciding with a \hi\
absorption. Figure~\ref{examples} (b) represents a far distance object
with no \hi\ SA located at the $^{13}$CO peak velocity, whereas
Figure~\ref{examples} (c) shows an example of an ambiguity, as overall
the $^{13}$CO peak lies in a \hi\ trough, but within the trough is a
spike, hence not providing any clear cut answer to the kinematic
distance ambiguity.

The majority (70\%) of the distance ambiguity cases had spectral
profiles indicating a clear near or far distance solution. Of the 94
sources, 33 of these are now determined to be at the near distance,
and 33 objects at the far distance. A futher 24 objects had line
features (described below) that were unable to provide a definite
answer to the distance ambiguity, and so were designated a probable
(best) solution. The distance ambiguities to 4 objects were unable to
be solved by using this method.

The results of the Gaussian fits and resulting kinematic distances for
all objects and any line-of-sight components can be found in Table
1. Direct overlays of all the $^{13}$CO 1-0 molecular line and \hi\
21cm data for each of the RMS sources can be found in
Appendix Figure~\ref{results}.

Many $^{13}$CO spectra show more than one emission component from
clouds at different distances along the line of sight, with an
increase in the number of objects towards the Galactic centre.. The
results also show that we can simultaneously solve the kinematic
distance ambiguity to many of these other clouds along the line of
sight. Of the other 114 component $^{13}$CO peaks 31 overlap with \hi\
SA, indicating near distance objects, and the lack of \hi\ SA for 57
components imply far distance objects. Again 22 components were unable
to provide a definite answer to the distance ambiguity and were
designated a probable (best) solution. Four components were left
undetermined.

The \hi\ SA technique is much easier to use in resolving multiple
line-of-sight component kinematic distance ambiguities than using the
H110$\alpha$ recombination line and H$_{2}$CO absorption technique
(such as Sewilo \textit{et al.} 2004). When using the RRL technique it
is not always possible to determine which H$_{2}$CO lines are being
absorbed against which \hii\ region if there are multiple \hii\
regions within the beam. For example, if multiple radio recombination
lines are detected towards a source, and only one component with a
distinct absorption, it is not possible to determine which \hii\
region is responsible for the absorption, and hence resolve the
distance ambiguity to these sources. Using the \hi\ SA technique a
direct comparison of both data sets is easily achieved without any
similar problems, even if multiple clouds may still be located within
the beam.

\subsection{Limitations due to complex line profiles}

Approximately 16\% of objects are either self-absorbed or have a
double-peaked profile at $^{13}$CO (eg. G299.5265+00.1478).  Structure
in the $^{13}$CO line makes it more difficult to match features in the
\hi\ spectrum.  Self absorption, while uncommon in $^{13}$CO, can
reduce the overall line strength making it weaker, whilst also
affecting matching up potential dips with the line profile when
overlaid onto \hi. This may push the result to either a near or far
distance depending on the peak of the stronger line. The overall line
profile was considered when overlaid onto the \hi\ data where cases
with suspected $^{13}$CO self absorption occurs and noted in the
results table.

A similar strategy to the self-absorbed lines was used when $^{13}$CO
emission at the off position occurred at the velocity of the $^{13}$CO
emission. These objects in the off beam are due to high Galactic
source crowding towards the inner plane of the Galaxy
(eg. G338.2459-00.6115). Although great care was taken to try and
reduce this effect whilst observing, some off beam objects still
remain. The overall line profile again was considered before any
decisions were made.

A few $^{13}$CO lines had saturated flat tops, or structured peak
profile such as G337.0642-01.1723, providing a unique problem. If we
were to consider the line profile as a whole, one could say it should
be at the the far distance or ambiguous. However, just considering the
spiked edge, this lies well within a \hi\ dip, and hence would be
located at the near distance. For objects such as this the distance
remains ambiguous. Similarly, $^{13}$CO peaks lying halfway down a
side of a \hi\ dip (e.g. G311.9799-00.9527) again provide similar
problems. The sources that remain ambiguous are noted in Appendix
Table A1.

Noisy \hi\ profiles are sometimes seen towards the upper or lower
edges ($b >\pm 1.3^\circ$) of a spectral cube due to more sparsely
sampled and weaker \hi\ data at these positions. This reduces the
quality and resolution of the \hi\ data such as with object
G340.2637-01.3236, and also complicates the comparison of the two
spectra. 

One interesting question is how big an absorption in \hi\ is required
to classify the object in question to be at a near or far
distance. For objects which do not have other features affecting the
classification of a near or far distance, a dip of $\sim$15-20\%\ of
the surrounding continuum intensity was required (shown by \citet{28})
to classify a near distance object. Components that had less \hi\
absorption remain ambiguous, however, were designated a probable near
or far solution as indicated by features explained in Appendix Table
A1.

\subsection{Further testing of the \hi\ SA method}

The likelihood of detecting \hi\ SA favours the geometry in
which a molecular cloud lies at the near kinematic distance. Since
\hi\ exists everywhere throughout the Galaxy, it emits at essentially
every allowed velocity. For clouds at the near kinematic distance,
there is an ample amount of \hi\ at the far distance and at the same
velocity against which a cold molecular foreground cloud can be
seen. If the same cloud were placed at the far distance, however,
there should be little or no background \hi, and the SA feature would
be absent. In addition, any weak SA for far clouds would be filled in
by emission from warm clouds in the foreground. Note object
G291.1881-00.2729 component 1, which is physically solved to be at the
far distance, yet shows very slight \hi\ SA features within its
spectrum for example.

\hi\ self-absorption may occur for clouds at the far kinematic
distance, from noncircular velocities due to random cloud motions or
streaming motions due to spiral density waves. This could change a
distant cloud's velocity enough to place it in front of a significant
column of gas which is at the same LSR velocity, but located behind
the cloud at the far kinematic distance.

This led us to testing a further sample of 10 objects which are
geometrically solved to be located outside the solar circle. Hence,
they have no distance ambiguity associated with them due to a
breakdown of the solutions to the kinematic distance ambiguity. These
should contain no \hi\ SA at the corresponding $^{13}$CO
velocity. Direct overlays of all the $^{13}$CO 1-0 molecular line and
\hi\ 21cm data for each of the geometrically solved MYSO candidates
can be found in Appendix Figure~\ref{outer}. For 7 of these objects
there is no \hi\ SA at the $^{13}$CO velocity indicating a clear far
distance solution as expected. However, for the other 3 objects,
G281.5961-00.1362, G282.2992-00.7763 and G306.1260-01.1371 there is a
slight dip in each of the \hi\ spectra at the $^{13}$CO velocity
position, indicating \hi\ being located at that velocity, behind the
object in question. This may be explained by looking at their location
within the Galaxy.  G281.5961-00.1362 and G282.2992-00.7763 are both
tangential along the line-of-sight to the Carina-Sagittarius arm, with
the arm extending far behind them. This could provide the streaming
motions required to put the \hi\ column of gas at the same LSR
velocity but located behind the cloud at the far kinematic
distance. G306.1260-01.1371 (and G282.2992-00.7763) however, is
located on the edge of the solar circle between two spiral arms, and
hence is only `just' solved geometrically and may have peculiar
velocities associated with it that could allow the solution to pass
over the solar circle (i.e. a v$_{LSR}$~$<$~0), and hence have a
kinematic distance ambiguity. Ignoring those objects for which a
peculiar velocity of 10 km s$^{-1}$ would allow the solution to pass
over the solar circle and have a kinematic distance ambiguity, the
method can be seen to be successful in 85\% (6 from 7) of cases.

A further confidence test was performed on the entire data set by
randomly shifting the $^{13}$CO spectrum along in velocity and
re-performing the \hi\ SA technique on the shifted data. This was
performed in order to check the random chance of a \hi\ dip lining up
with a $^{13}$CO component. The $^{13}$CO spectra were moved by both
$\pm$10 km s$^{-1}$ and replotted over the \hi\ line
data. Approximately 80\% of the resultant matches implied a far
distance i.e. no \hi\ SA, with only 10\% of the $^{13}$CO components
randomly coinciding with a \hi\ dip implying a near distance. A
further 10\% of components were unable to be determined, again due to
features mentioned previously. This implys an 80\% success rate in
determining the correct designation of whether an object is at the
near or far distance, although the numbers tested mean this rate is
uncertain by 14\%.

\subsection{Tracing Galactic structure}

As a check when dealing with distance ambiguities related to massive
star formation, it is natural to ask whether the RMS survey sources
studied follow any Galactic spiral arm structure, as it is well known
that young massive stars are good tracers of Galactic spiral structure
in galaxies \citep{26}.

\begin{figure}
\includegraphics[width=80mm]{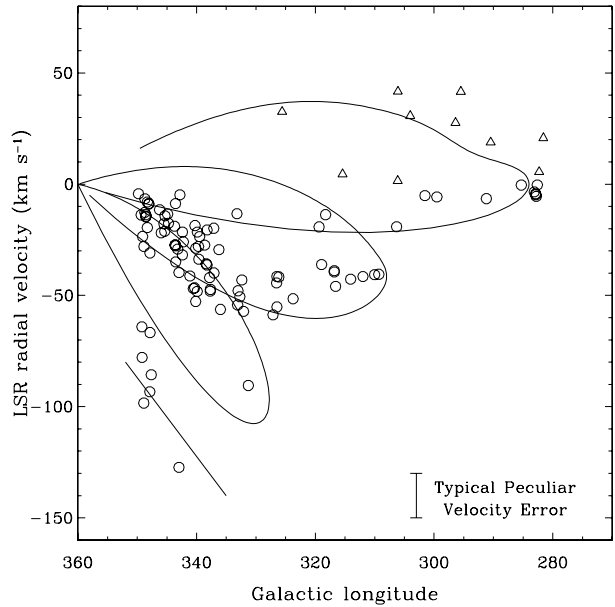}
 \caption{Longitude-velocity plot showing the distribution of our sources (open circles) in velocity space (i.e. no ambiguity). Open triangles represent the 10 sources which were studied with no distance ambiguity. Over-plotted are the Galactic arms from the model of \citet{16}. The three Galactic arms shown are from upper to lower: Carina-Sagittarius, Crux-Scutum and Norma arm. The short straight line represents the 3 kpc expanding arm. A typical peculiar velocity error $\pm$10 km s$^{-1}$ is shown, allowing many of the sources to be positioned on two Galactic arms towards the inner Galaxy.}
  \label{rotation}
\end{figure}

Figures ~\ref{galaxy} and ~\ref{rotation} help in trying to answer
this question. Figure ~\ref{galaxy} shows the projected location in
the Galactic plane of all our sources after solving the distance
ambiguity problem. In this figure only the 4th Galactic quadrant is
shown, with the Sun's location at the origin, and the Galactic centre
at (0, 8.5) kpc. Figure ~\ref{galaxy} shows only those points for
which the distance ambiguity can be reasonably resolved. Distance
errors relating to peculiar velocities of $\pm$ 10 km s$^{-1}$ are
also shown. Although deviations from circular motion of up to $\pm$ 15
km s$^{-1}$ can easily be found within the Galaxy \citep{27}, these
are generally more localised along the spiral arms.

Crowding towards the inner Galaxy can clearly be seen as expected,
especially with the near distance sources, however no sharp or
well-defined spiral structure can be easily seen. Objects at the far
distance tend to lie on or near spiral arms, however, this is not
always the case. Nearby objects towards the inner Galaxy tend to be
much worse, and when taking into account potential peculiar
velocities, some could lie on either of two Galactic arms. Also those
inner Galactic sources with $l > 340^{\circ}$ give more unreliable
distances due to radial velocities being dominated by peculiar motions
rather circular velocities.

The longitude-velocity (\textit{l-v}) plot shown in Figure
~\ref{rotation} is an attempt to reveal the spiral structure using
only observed quantities. The positions of the sources and components
are independent of any Galactic rotation model, hence show no
ambiguity. The spiral arm centroids are again plotted from the updated
model of \citet{16}. The obvious location crowding can again be seen
towards the inner Galaxy as many of the Galactic arms overlap in the
velocity domain. The tangent points of the outer two Galactic arms can
also be picked out. Although no major Galactic structure can be
ascertained from such a small sample, one interesting feature clearly
seen is the 3 kpc expanding arm. These sources are located towards the
inner Galaxy by the Norma arm of the Galaxy running down towards a
Galactic longitude of $340^\circ$ (indicated by a short line),
matching the 3 kpc expanding arm from the large-scale CO surveys of
\citet{18}. This could also account for the group of sources in Figure
~\ref{galaxy} located inside the Norma arm towards the Galactic
centre.

\section{Conclusions}

We used $^{13}$CO 1-0 and N$_{2}$H$^{+}$ observations combined with
SGPS \hi\ data to resolve the kinematic distance ambiguity problem
towards 94 RMS survey candidate MYSOs. The main motivation for this
was to obtain accurate distances to potential MYSOs within the inner
Galaxy, whilst also testing whether the \hi\ SA technique is valid on
our initial sample of sources. All sources studied had $^{13}$CO 1-0
spectral line or N$_{2}$H$^{+}$ detections, with just over 55\%\
containing multiple velocity components along the line-of-sight in
$^{13}$CO.

The majority (66) of the 94 RMS survey MYSO candidates with a distance
ambiguity were resolved using this method, with a further 24 objects
having features that were unable to provide a definite answer to the
distance ambiguity.  Other line-of-sight velocity components present
in the spectra were also solved simultaneously by this method.

The sources with resolved distance ambiguities were plotted on
Galactic position-position and \textit{l-v} plots to determine if
spiral arm patterns could be clearly discerned. No obvious Galactic
pattern could be seen, likely due to such a small sample size. Once
typical data scatter introduced by peculiar velocities is allowed for,
the source distance scatter can be the separation of the spiral arms
themselves. The positions of the sources in the velocity domain
provide a slightly better picture with the majority lying near loci of
spiral arms, including a small group of sources following the 3 kpc
expanding arm.

Obtaining accurate distances to our RMS survey MYSO candidates is
crucial in helping to determine the physical properties
(e.g. luminosities) of these potential massive young stellar objects
and the surrounding star formation regions. Whilst this is a
relatively unstudied quadrant of the Galaxy, with respect to distance
ambiguities, the completion of the SGPS now gives us access to \hi\
data in the entire southern hemisphere, allowing now for a much more
thorough test of the \hi\ SA technique. With the majority of the IGPS
\hi\ data being readily available, this provides a wealth of
information, ripe for large scale (kinematic) surveys such as our own
RMS survey.

\section*{Acknowledgements}

We would like to thank Naomi McClure-Griffiths for her time and
patience in providing all the \hi\ data cubes which we required. We
would also like to thank the anonymous referee whose comments provided
a useful change of the structure and reasoning of the paper.

\appendix
\section[]{Results}

We plot direct overlays of the $^{13}$CO 1-0 line and \hi\ data for
all objects with a distance ambiguity in
Figure~\ref{results}. 

\begin{figure*}
\includegraphics[width=170mm]{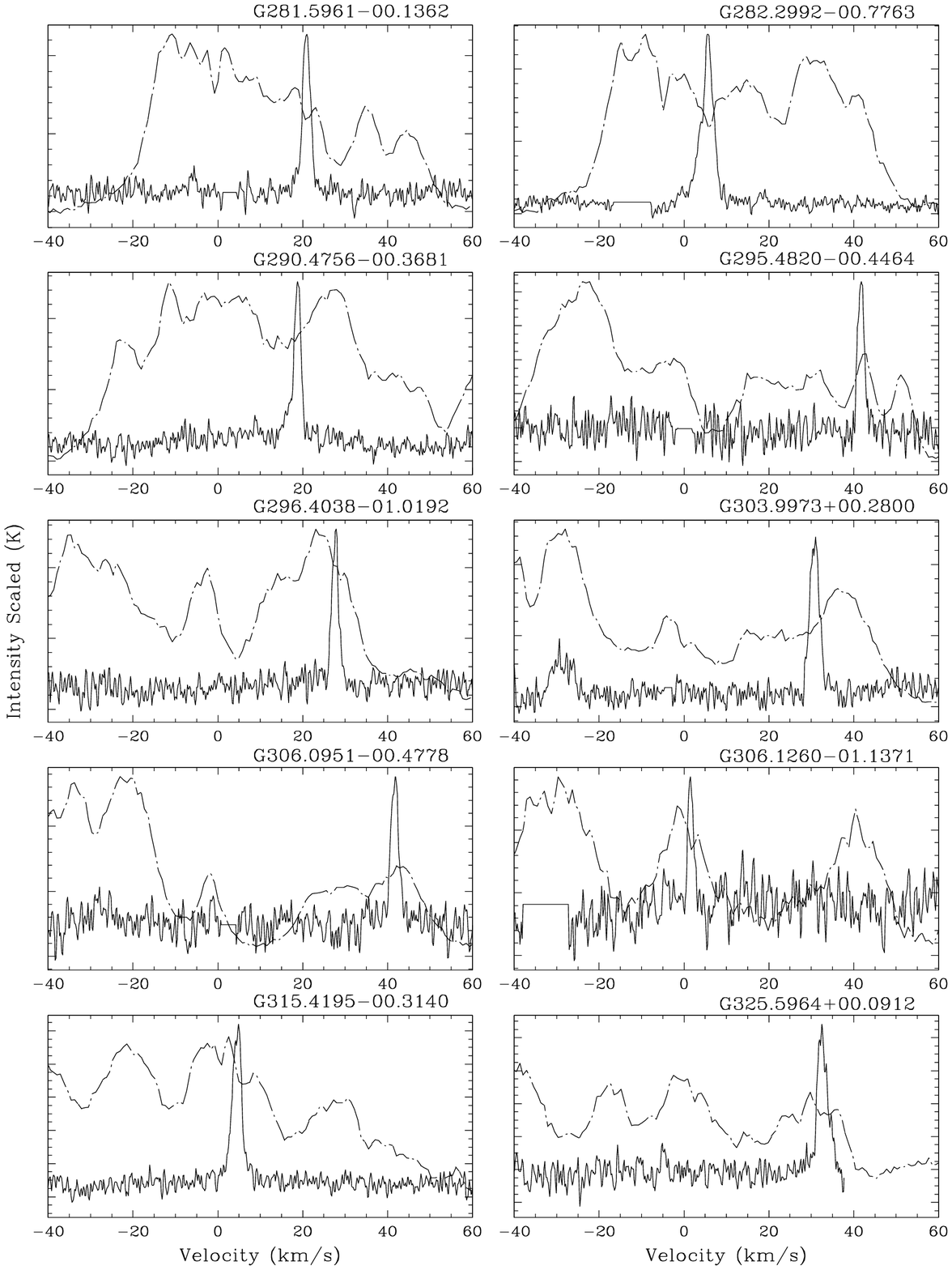}
 \caption{Plots showing the direct overlays of the $^{13}$CO 1-0 line and \hi\ data on objects that are geometrically solved to be outside the solar circle, and hence have no distance ambiguity associated with them. The dashed line spectra represents the \hi\ through the Galaxy, whereas the solid line is the $^{13}$CO spectra of the \textit{MSX} object in question.}
 \label{outer}
\end{figure*}

Overlays of all the $^{13}$CO 1-0 molecular line and \hi\ 21cm data
for each of the geometrically solved MYSO candidates can be found in
Figure~\ref{outer}. 

Tabulated results of all calculated parameters for all objects with a
distance ambiguity can be found in Table A1. All calculated parameters
for the 10 objects which are geometrically solved to be at the far
distance can be found in Table A2.

\begin{figure*}
\includegraphics[width=170mm]{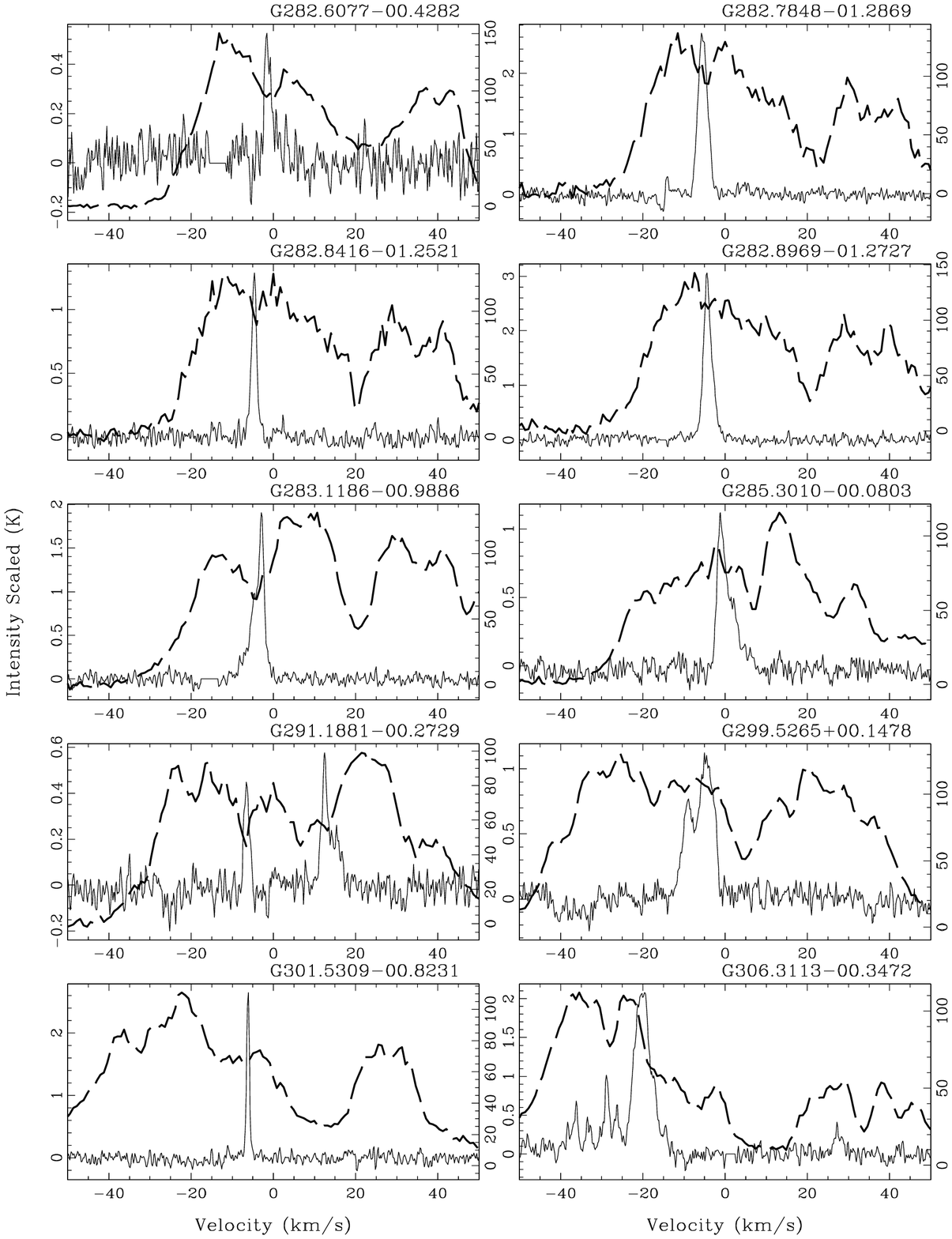}
 \caption{Plots showing the direct overlays of the $^{13}$CO 1-0 line and \hi\ data. The bold dashed line spectra represents the \hi\ through the Galaxy, whereas the solid line is the $^{13}$CO spectra of the \textit{MSX} object in question with the peak scaled to that of \hi. An asterisk after the \textit{MSX} name represents a N$_{2}$H$^{+}$ line being used instead of $^{13}$CO line. The left and right hand axes represent $^{13}$CO and \hi\ intensities respectively.}
  \label{results}
\end{figure*}

\begin{figure*}
\includegraphics[width=175mm]{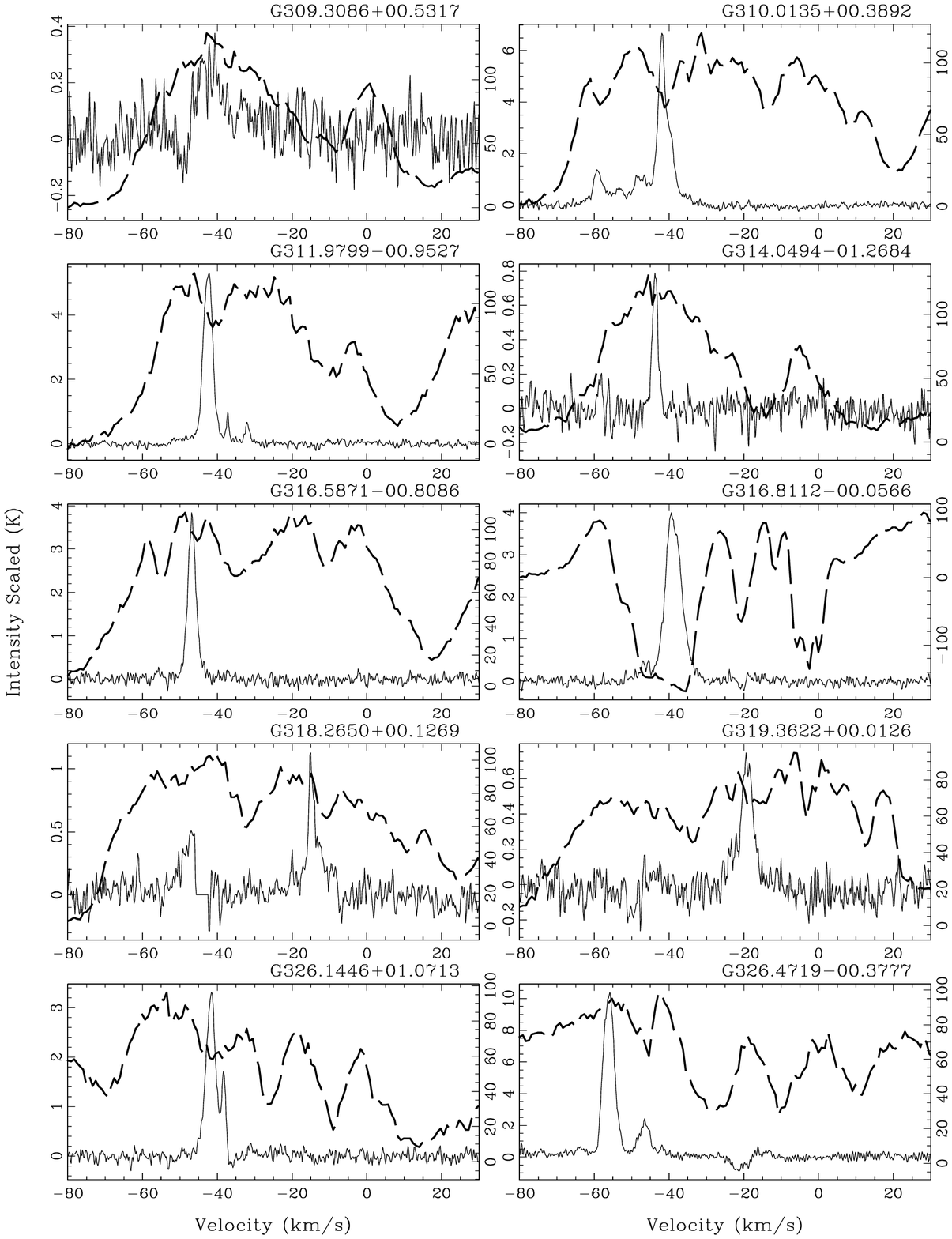}
 \contcaption{}
\end{figure*}

\begin{figure*}
\includegraphics[width=175mm]{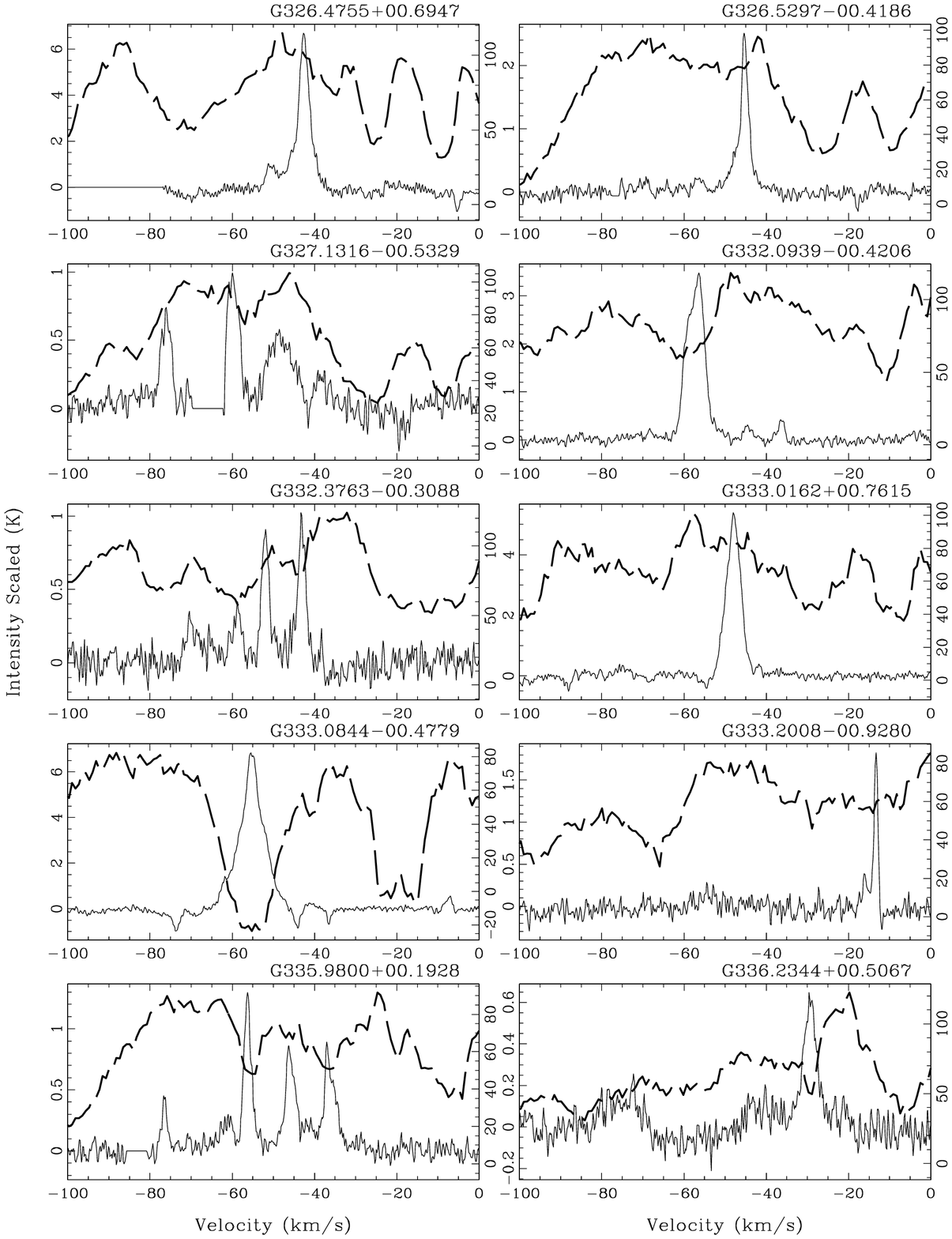}
 \contcaption{}
\end{figure*}

\begin{figure*}
\includegraphics[width=175mm]{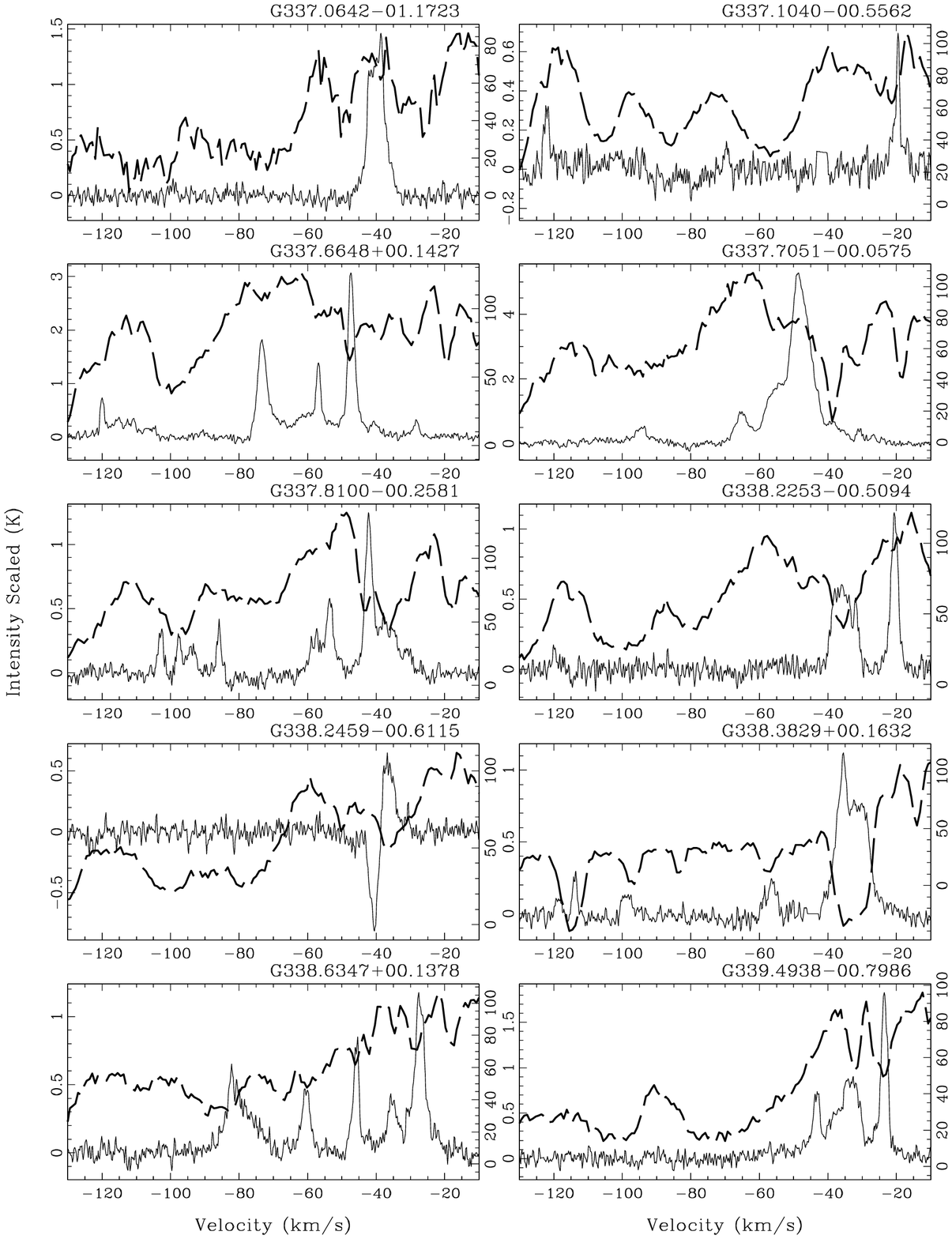}
 \contcaption{}
\end{figure*}

\begin{figure*}
\includegraphics[width=175mm]{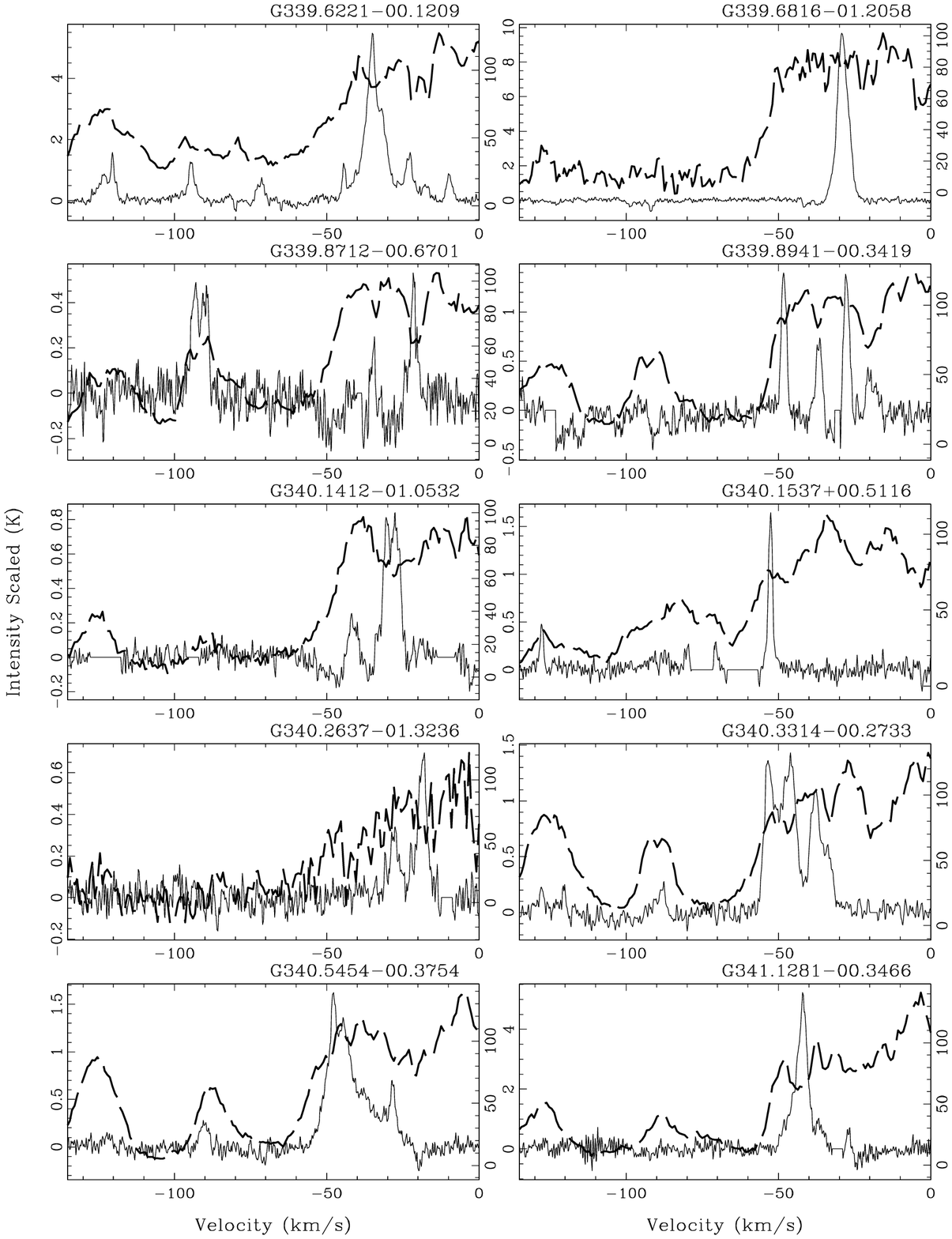}
 \contcaption{}
\end{figure*}

\begin{figure*}
\includegraphics[width=175mm]{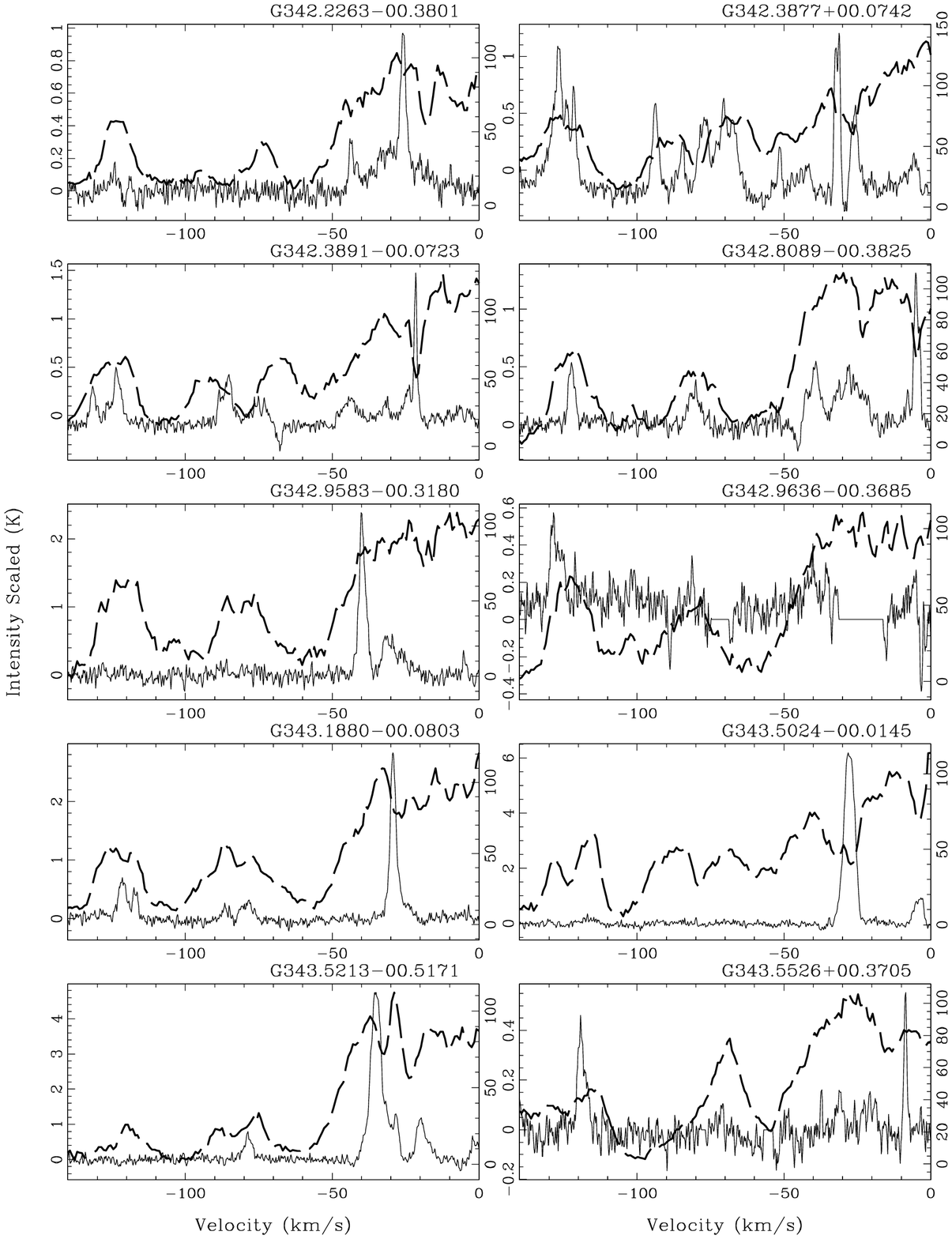}
 \contcaption{}
\end{figure*}

\begin{figure*}
\includegraphics[width=175mm]{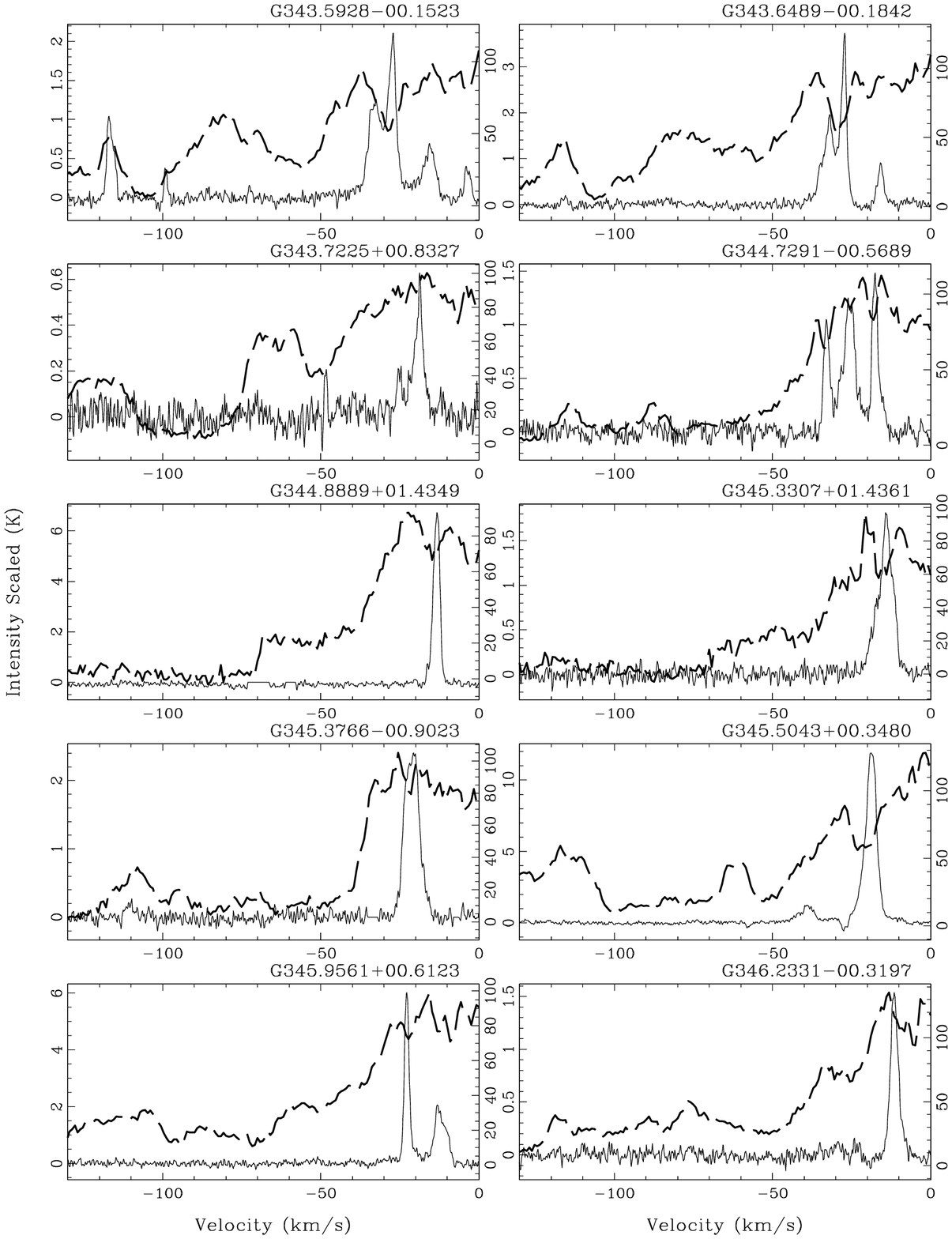}
 \contcaption{}
\end{figure*}

\begin{figure*}
\includegraphics[width=175mm]{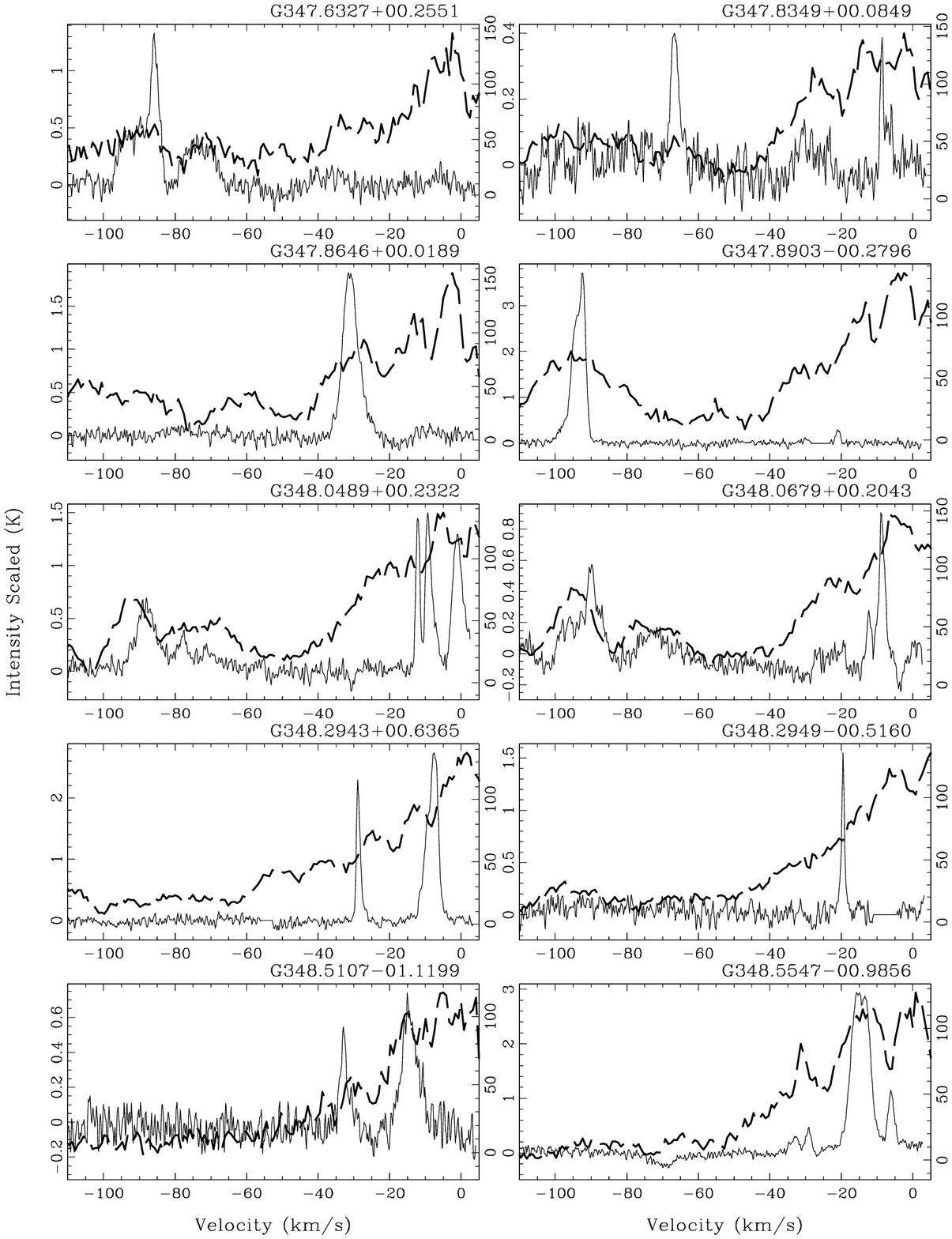}
 \contcaption{}
\end{figure*}

\begin{figure*}
\includegraphics[width=175mm]{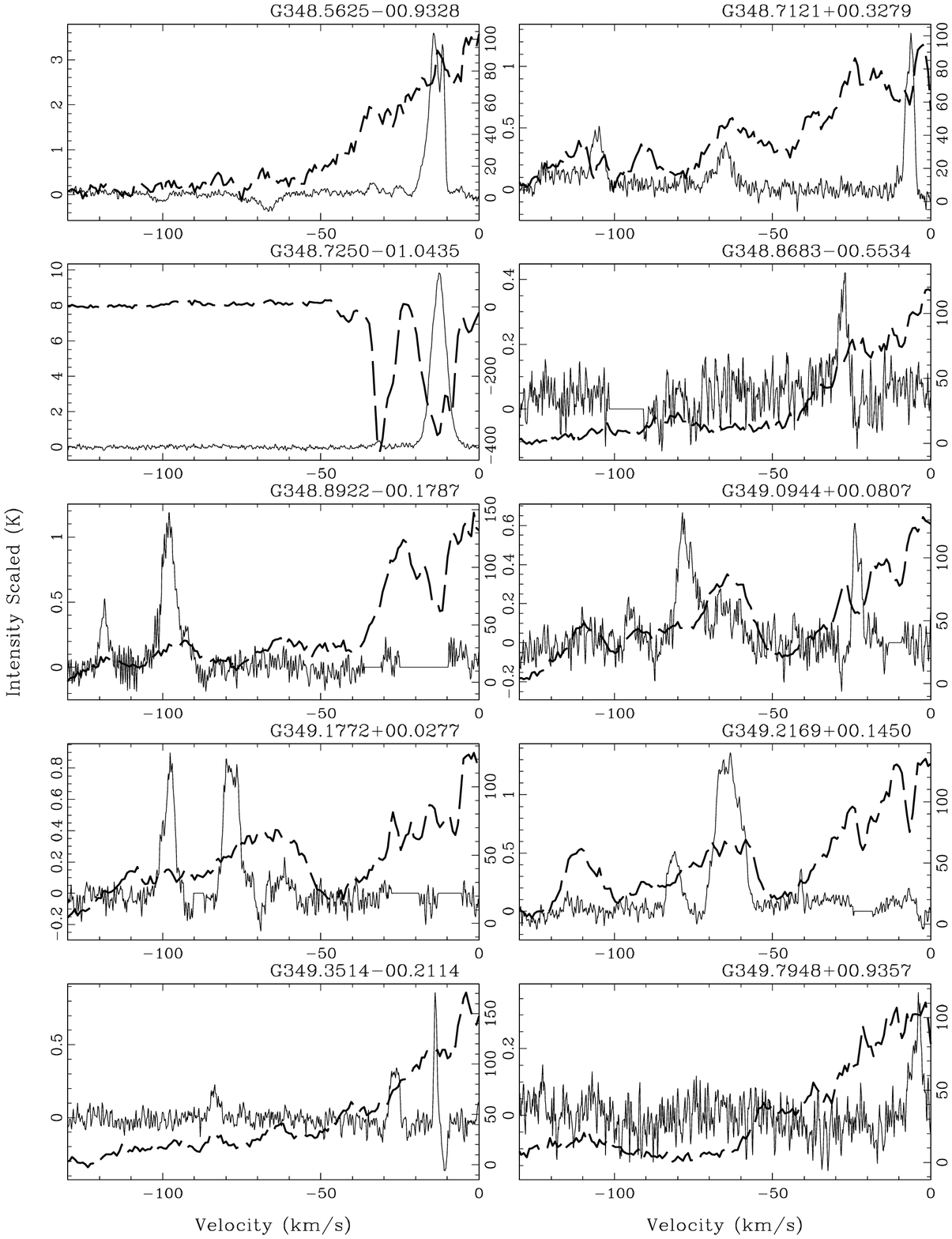}
 \contcaption{}
\end{figure*}

\begin{figure*}
\includegraphics[width=175mm]{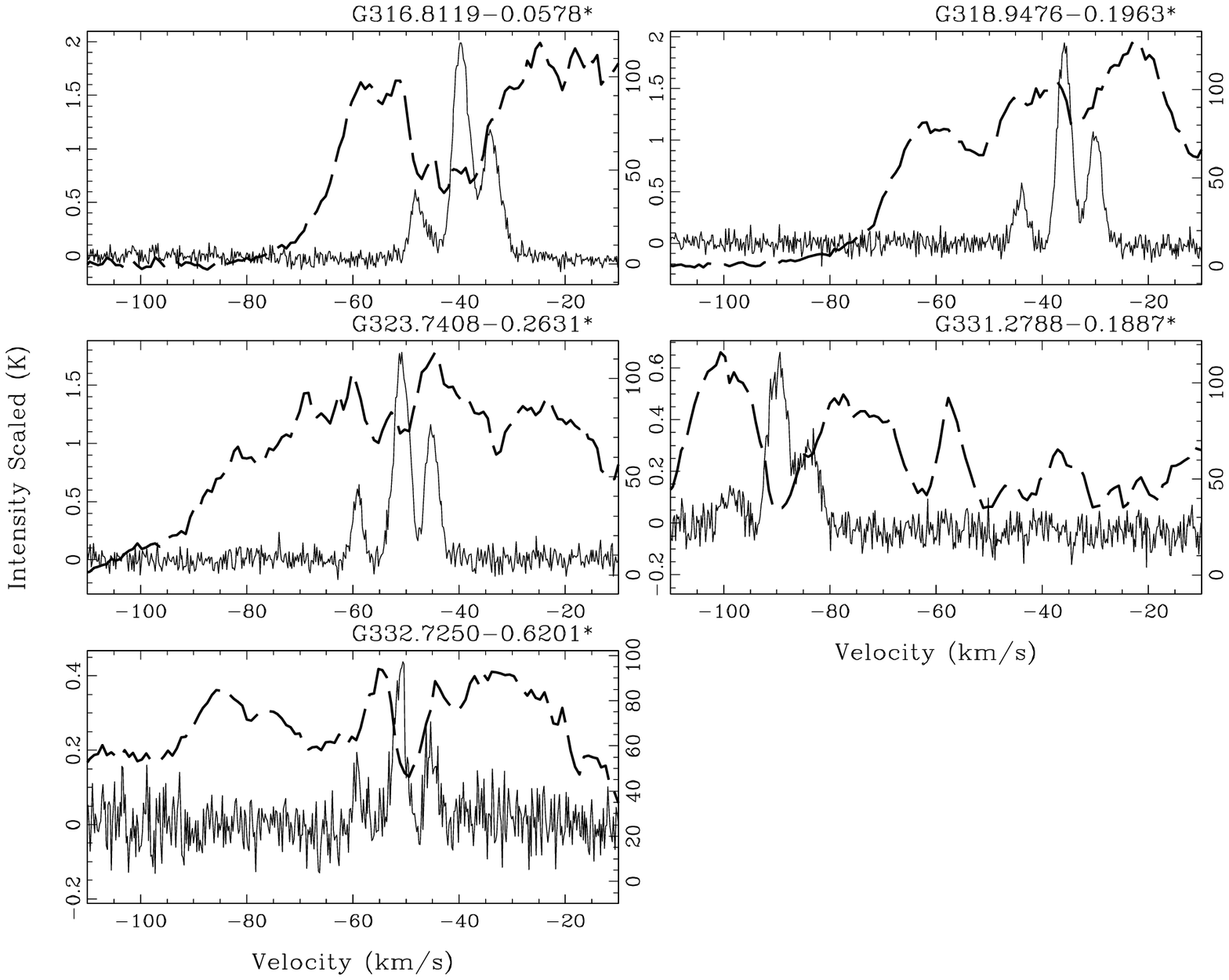}
\vspace{9cm}
 \contcaption{}
\end{figure*}

\onecolumn

\begin{center}

\tablefirsthead{ 
 \multicolumn{9}{c}{Table A1. $^{13}$CO 1-0 Kinematic parameters of \textit{MSX} sources with a distance ambiguity}\\
  \hline
     \hline
     \multicolumn{1}{c}{Source Name} &  & \multicolumn{1}{c}{V$_{LSR}$} & \multicolumn{1}{c}{D$_{near}$} & \multicolumn{1}{c}{D$_{far}$} & \multicolumn{1}{c}{z} & \multicolumn{1}{c}{R$_{gc}$} & Adopted & Notes \\ 
     & & (km s$^{-1}$) & (kpc) & (kpc) & (pc) & (kpc) & Distance & \\
     \hline}

\tablehead{ 
  \multicolumn{9}{c}{\small {Table A1 - Continued.}} \\
  \hline
  \hline
     \multicolumn{1}{c}{Source Name} &  & \multicolumn{1}{c}{V$_{LSR}$} & \multicolumn{1}{c}{D$_{near}$} & \multicolumn{1}{c}{D$_{far}$} & \multicolumn{1}{c}{z} & \multicolumn{1}{c}{R$_{gc}$} & Adopted & Notes \\ 
     & & (km s$^{-1}$) & (kpc) & (kpc) & (pc) & (kpc) & Distance & \\
     \hline}

\tablelasttail{\hline}

\small{
\begin{mpsupertabular}{llccccccl}
G282.6077-00.4282 &   & -0.3   & \bf{0.1}  (1.4)&	3.6  &	      -0.7    & 8.5  & N  & \\
G282.7848-01.2869 &   & -5.5   & 1.8            &	1.9  &	      -       & 8.3  & ?  & ambiguous \\
G282.8416-01.2521 &   & -4.7   & \bf{1.2}  (1.9)&	2.6  &	      -26.2   & 8.3  & N  & \\
G282.8969-01.2727 &   & -4.3   & 1.0            &\bf{2.8} (1.8)&      -62.2   &	8.3  & F  & \\
G283.1186-00.9886 &   & -3.5   & \bf{0.7}  (1.6)&	3.1       &   -12.1   &	8.4  & N? & \\
G285.3010-00.0803 &   & -0.4   & 0.1            &\bf{4.4} (1.2)&      -6.2    &	8.5  & F  & \\
G291.1881-00.2729 & 1 & 13.0   & -              &	7.5  (0.9)&   -35.7   &	9.1  & -  & n/a \\
                  & 2 & -6.5   & \bf{0.8}  (1.2)&	5.3       &   -3.8    &	8.2  & N  & \\
G299.5265+00.1478 &   & -5.7   & 0.5            &\bf{7.9} (0.9)&      20.4    &	8.3  & F  & SA/off-beam? \\
G301.5309-00.8231 &   & -5.1   & 0.4            &\bf{8.4} (0.8)&      -120.7  &	8.3  & F  & \\
G306.3113-00.3472 & 1 & -19.1  & 1.6            &\bf{8.5} (0.8)&      -51.5   &	7.7  & F  & \\
                  & 2 & -27.2  & \bf{2.3}  (0.9)&	7.8       &   -13.9   &	7.4  & N  & \\
G309.3086+00.5317 &   & -40.5  & 3.4            &\bf{7.4} (0.9)&      68.7    &	6.9  & F  & off-beam \\
G310.0135+00.3892 & 1 & -40.7  & \bf{3.3}  (0.9)&	7.6       &   22.4    &	6.9  & N  & \\
                  & 2 & -58.7  & -              &  -           &      -       &	6.3  & -  & Unphysical V$_{LSR}$\\
                  & 3 & -47.5  & 4.2            &\bf{6.7} (1.1)&      45.5    &	6.6  & F  & \\
G311.9799-00.9527 & 1 & -41.5  & \bf{2.2}  (0.8)&	8.1       &   -36.6   &	6.8  & N? & \\
                  & 2 & -36.3  & 2.8            &\bf{8.6} (0.7)&      -143.0  &	7.0  & F  & \\
                  & 3 & -31.0  & \bf{2.4}  (0.7)&	9.0       &   -39.9   &	7.1  & N? & \\
G314.0494-01.2684 &   & -42.7  & 3.2       (0.7)&	8.6 (0.7) &   -       &	6.7  & ?  & ambiguous \\
G316.5871-00.8086 &   & -45.9  & \bf{3.3}  (0.7)&	9.1       &   -46.6   &	6.5  & N? & \\
G316.8112-00.0566 &   & -38.9  & \bf{2.8}  (0.6)&	9.6       &   -2.8    &	6.8  & N  & \\
G318.2650+00.1269 & 1 & -13.7  & 1.0            &\bf{11.7}(0.7)&      25.9    &	7.8  & F  & \\
                  & 2 & -46.4  & 3.2            &\bf{9.5} (0.6)&      21.0    &	6.5  & F? & off-beam \\
G319.3622+00.0126 &   & -19.2  & \bf{1.4}  (0.7)&	11.5      &   0.3     &	7.5  & N? & \\
G326.1446+01.0713 & 1 & -41.7  & \bf{2.8}  (0.5)&	11.3      &   52.4    &	6.3  & N? & SA/off-beam? \\
                  & 2 & -38.3  & \bf{2.6}  (0.6)&	11.5      &   48.6    &	6.5  & N? & \\
G326.4719-00.3777 & 1 & -55.2  & 3.6            &\bf{10.6}(0.5)&      -69.9   &	5.9  & F  & \\
                  & 2 & -45.7  & \bf{3.1}  (0.5)&	11.1      &   -20.4   &	6.2  & N  & \\
G326.4755+00.6947 &   & -41.5  & 2.8            &\bf{11.3}(0.5)&      137.0   &	6.3  & F  & \\
G326.5297-00.4186 &   & -44.4  & 3.0            &\bf{11.2}(0.5)&      -81.8   &	6.2  & F  & \\
G327.1316-00.5329 & 1 & -47.4  & \bf{3.2}  (0.5)&	11.1      &   -29.8   &	6.1  & N? & strongest? \\
                  & 2 & -58.7  & 3.8            &\bf{10.5}(0.5)&      -97.7   &	5.7  & F  & off-beam \\
                  & 3 & -75.2  & 4.7            &\bf{9.6} (0.5)&      -89.3   &	5.2  & F  & \\
G332.0939-00.4206 & 1 & -57.2  & \bf{3.8}  (0.5)&	11.3      &   -27.9   &	5.5  & N? & \\
                  & 2 & -36.3  & 2.6            &\bf{12.4}(0.6)&      -91.0   &	6.3  & F  & \\
G332.3763-00.3088 & 1 & -43.1  & \bf{3.0}  (0.5)&	12.0      &   -16.2   &	6.0  & N  & \\
                  & 2 & -51.9  & 3.5            &\bf{11.6}(0.5)&      -62.5   &	5.6  & F  & \\
                  & 3 & -58.8  & \bf{3.9}  (0.5)&	11.2      &   -21.0   &	5.4  & N  & \\
G333.0162+00.7615 &   & -47.9  & 3.3            &\bf{11.8}(0.5)&      156.8   &	5.7  & F  & \\
G333.0844-00.4779 & 1 & -54.3  & \bf{3.6}  (0.5)&	11.5      &   -30.0   &	5.5  & N  & \\
                  & 2 & -6.4   & 0.6            &\bf{14.6}(0.9)&      -121.8  &	8.0  & F  & \\
G333.2008-00.9280 &   & -13.3  & 1.1            &\bf{14.0}(0.8)&      -226.8  &	7.5  & F? & \\
G335.9800+00.1928 & 1 & -56.3  & \bf{3.9}  (0.5)&	11.7      &   13.1    &	5.2  & N  & \\
                  & 2 & -45.9  & \bf{3.3}  (0.5)&	12.2      &   11.1    &	5.6  & N? & \\
                  & 3 & -36.4  & \bf{2.8}  (0.6)&	12.7      &   9.4     &	6.0  & N  & \\
                  & 4 & -76.5  & 4.8            &\bf{10.8}(0.4)&      36.3    &	4.6  & F  & \\
G336.2344+00.5067 &   & -29.4  & \bf{2.4}  (0.6)&	13.2      &   21.2    &	6.4  & N  & \\
G337.0642-01.1723 &   & -39.9  & 3.1       (0.6)&	12.6 (0.6)&   -       &	5.8  & ?  & ambiguous \\
G337.1040-00.5562 & 1 & -19.8  & \bf{1.8}  (0.8)&	13.9      &   -17.5   &	6.9  & N? & \\
                  & 2 & -122.0 & \bf{6.7}  (0.4)&	9.0       &   -65.0   &	3.5  & N  & \\
G337.6648+00.1427 & 1 & -47.3  & \bf{3.5}  (0.5)&	12.2      &   8.7     &	5.4  & N  & \\
                  & 2 & -73.2  & \bf{4.7}  (0.4)&	11.0      &   11.7    &	4.5  & N  & \\
                  & 3 & -57.0  & 4.0            &\bf{11.8}(0.4)&      29.4    &	5.1  & F  & \\
                  & 4 & -119.9 & 6.5            &\bf{9.2} (0.4)&      22.9    &	3.5  & F  & \\
                  & 5 & -28.4  & \bf{2.4}  (0.7)&	13.3      &   6.0     &	6.4  & N  & \\
G337.7051-00.0575 & 1 & -48.1  & 3.5            &\bf{12.2}(0.5)&      -12.2   &	5.4  & F  & structure \\
                  & 2 & -56.4  & 3.9            &\bf{11.8}(0.4)&      -11.8   &	5.1  & F  & \\
                  & 3 & -65.0  & \bf{4.3}  (0.4)&	11.4      &   -4.3    &	4.8  & N? & \\
G337.8100-00.2581 & 1 & -42.0  & \bf{3.2}  (0.5)&	12.5      &   -14.4   &	5.7  & N  & \\
                  & 2 & -54.5  & \bf{3.9}  (0.5)&	11.9      &   -17.6   &	5.1  & N? & \\
                  & 3 & -85.9  & 5.2            &\bf{10.5}(0.3)&      -47.3   &	4.2  & F  & \\
                  & 4 & -102.9 & 5.8            &\bf{9.9} (0.3)&      -44.6   &	3.8  & F  & \\
                  & 5 & -95.4  & 5.6            &\bf{10.2}(0.3)&      -45.9   &	4.0  & F  & SA/off-beam? \\
G338.2253-00.5094 & 1 & -35.6  & \bf{2.9}  (0.6)&	12.9      &   -25.8   &	5.9  & N  & strongest? \\
                  & 2 & -20.6  & \bf{1.9}  (0.8)&	13.9      &   -16.9   &	6.8  & N  & \\
G338.2459-00.6115 &   & -36.3  & \bf{2.9}  (0.6)&	12.9      &   -31.0   &	5.9  & N  & off-beam \\
G338.3829+00.1632 & 1 & -35.8  & \bf{2.9}  (0.6)&	12.9      &   8.3     &	5.9  & N  & structure \\
                  & 2 & -113.8 & \bf{6.3}  (0.3)&	9.5       &   17.9    &	3.5  & N  & \\
                  & 3 & -56.6  & \bf{4.0}  (0.4)&	11.8      &   11.4    &	5.0  & N  & \\
                  & 4 & -97.9  & \bf{5.7}  (0.3)&	10.1      &   16.2    &	3.9  & N  & \\
G338.6347+00.1378 & 1 & -27.4  & \bf{2.4}  (0.7)&	13.4      &   5.8     &	6.3  & N  & \\
                  & 2 & -45.9  & \bf{3.5}  (0.5)&	12.3      &   8.4     &	5.4  & N  & \\
                  & 3 & -80.5  & \bf{5.0}  (0.3)&	10.8      &   12.0    &	4.2  & N? & strongest? \\
                  & 4 & -60.5  & \bf{4.2}  (0.4)&	11.6      &   10.1    &	4.8  & N  & \\
                  & 5 & -35.0  & \bf{2.9}  (0.6)&	13.0      &   7.0     &	5.9  & N  & \\
G339.4938-00.7986 & 1 & -34.0  & \bf{2.9}  (0.6)&	13.1      &   -40.4   &	5.9  & N  & strongest? \\
                  & 2 & -23.6  & \bf{2.2}  (0.7)&	13.8      &   -30.7   &	6.5  & N  & \\
                  & 3 & -43.4  & 3.4            &\bf{12.5}(0.5)&      -174.2  &	5.4  & F  & \\
G339.6221-00.1209 & 1 & -33.7  & \bf{2.9}  (0.6)&	13.1      &   -6.1    &	5.9  & N  & \\
                  & 2 & -22.1  & \bf{2.1}  (0.8)&	13.9      &   -4.4    &	6.6  & N  & \\
                  & 3 & -93.5  & 5.5            &\bf{10.4}(0.3)&      -21.9   &	3.8  & F  & \\
                  & 4 & -120.8 & 6.5            &\bf{9.4} (0.3)&      -19.8   &	3.3  & F  & \\
                  & 5 & -43.3  & 3.4            &\bf{12.5}(0.5)&      -26.4   &	5.4  & F  & \\
                  & 6 & -8.8   & 0.9            &\bf{15.0}(1.0)&      -31.7   &	7.6  & F  & \\
                  & 7 & -70.7  & 4.7            &\bf{11.3}(0.4)&      -23.8   &	4.4  & F  & \\
G339.6816-01.2058 &   & -28.0  & 2.5            &\bf{13.4}(0.7)&      -282.0  &	6.2  & F? & \\
G339.8712-00.6701 & 1 & -91.8  & 5.5            &\bf{10.5}(0.3)&      -122.8  &	3.8  & F? & strongest? \\
                  & 2 & -21.5  & \bf{2.0}  (0.8)&	13.9      &   -23.4   &	6.6  & N  & \\
                  & 3 & -34.7  & \bf{2.9}  (0.6)&	13.0      &   -33.9   &	5.8  & N  & \\
G339.8941-00.3419 & 1 & -48.2  & \bf{3.7}  (0.5)&	12.3      &   -22.1   &	5.2  & N  & \\
                  & 2 & -27.6  & 2.5            &\bf{13.5}(0.7)&      -80.6   &	6.2  & F? & \\
                  & 3 & -36.7  & \bf{3.1}  (0.6)&	12.9      &   -18.5   &	5.7  & N  & \\
G340.1412-01.0532 & 1 & -28.8  & \bf{2.6}  (0.7)&	13.4      &   -47.8   &	6.1  & N? & SA/off-beam? \\
                  & 2 & -41.4  & 3.3            &\bf{12.6}(0.5)&      -231.6  &	5.5  & F  & \\
G340.1537+00.5116 & 1 & -52.7  & 3.9            &\bf{12.1}(0.5)&      108.0   &	5.0  & F? & \\
                  & 2 & -127.8 & 6.8            &\bf{9.2} (0.3)&      82.1    &	3.1  & F  & \\
G340.2637-01.3236 & 1 & -18.6  & \bf{1.8}  (0.8)&	14.2      &   -41.6   &	6.8  & N? & \\
                  & 2 & -27.8  & 2.5            &\bf{13.5}(0.7)&      -311.9  &	6.2  & F? & \\
G340.3314-00.2733 & 1 & -46.6  & \bf{3.6}  (0.5)&	12.4      &   -17.2   &	5.2  & N  & Double?/SA \\
                  & 2 & -53.0  & 4.0            &\bf{12.0}(0.5)&      -57.2   &	5.0  & F  & \\
                  & 3 & -37.2  & 3.1            &\bf{12.9}(0.6)&      -61.5   &	5.7  & F  & \\
G340.5454-00.3754 & 1 & -46.9  & 3.7            &\bf{12.4}(0.5)&      -81.2   &	5.2  & F  & structure \\
                  & 2 & -28.2  & \bf{2.6}  (0.7)&	13.5      &   -17.0   &	6.1  & N? & \\
                  & 3 & -90.3  & 5.5            &\bf{10.6}(0.3)&      -69.5   &	3.8  & F  & \\
G341.1281-00.3466 &   & -41.3  & \bf{3.4}  (0.6)&	12.7      &   -20.6   &	5.4  & N  & \\
G342.2263-00.3801 & 1 & -25.9  & \bf{2.5}  (0.8)&	13.7      &   -16.6   &	6.1  & N? & \\
                  & 2 & -43.6  & \bf{3.6}  (0.5)&	12.5      &   -23.9   &	5.2  & N  & \\
G342.3877+00.0742 & 1 & -125.8 & 6.7            &\bf{9.5} (0.3)&      12.3    &	2.9  & F  & strongest? \\
                  & 2 & -31.8  & \bf{3.0}  (0.7)&	13.2      &   3.9     &	5.8  & N? & \\
                  & 3 & -25.9  & \bf{2.5}  (0.8)&	13.7      &   3.2     &	6.1  & N  & \\
                  & 4 & -93.8  & 5.7            &\bf{10.5}(0.3)&      13.6    &	3.5  & F  & \\
                  & 5 & -69.5  & 4.8            &\bf{11.4}(0.4)&      14.8    &	4.2  & F  & \\
                  & 6 & -77.5  & \bf{5.1}  (0.3)&	11.1      &   6.6     &	3.9  & N? & \\
                  & 7 & -84.8  & 5.4            &\bf{10.8}(0.3)&      14.0    &	3.7  & F  & \\
                  & 8 & -51.4  & 4.1            &\bf{12.1}(0.5)&      15.7    &	4.8  & F  & \\
                  & 9 & -6.3   & \bf{0.8}  (1.2)&	15.4      &   1.0     &	7.8  & N? & \\
G342.3891-00.0723 & 1 & -21.5  & \bf{2.2}  (0.8)&	14.0      &   -2.8    &	6.4  & N  & \\
                  & 2 & -85.8  & 5.4            &\bf{10.8}(0.3)&      -13.6   &	3.7  & F  & \\
                  & 3 & -123.5 & 6.6            &\bf{9.6} (0.3)&      -12.1   &	3.0  & F  & \\
                  & 4 & -131.3 & 6.9            &\bf{9.4} (0.3)&      -11.9   &	2.9  & F  & \\
G342.8089-00.3825 & 1 & -122.3 & 6.6            &\bf{9.7} (0.2)&      -64.8   &	3.0  & F  & \\
                  & 2 & -4.8   & \bf{0.6}  (1.3)&	15.6      &   -4.0    &	7.9  & N  & \\
                  & 3 & -39.6  & 3.5            &\bf{12.8}(0.6)&      -85.5   &	5.3  & F  & \\
                  & 4 & -27.5  & 2.7       (0.7)&	13.5 (0.7)&   -       &	6.0  & ?  & ambiguous \\
                  & 5 & -80.6  & 5.3            &\bf{11.0}(0.3)&      -73.4   &	3.8  & F  & \\
G342.9583-00.3180 & 1 & -39.7  & 3.5            &\bf{12.8}(0.6)&      -71.0   &	5.3  & F  & \\
                  & 2 & -30.0  & \bf{2.9}  (0.7)&	13.4 (0.7)&   -16.1   &	5.8  & N? & \\
G342.9636-00.3685 &   & -127.3 & 6.7            &\bf{9.5} (0.2)&      -61.1   &	2.9  & F  & \\
G343.1880-00.0803 & 1 & -29.2  & \bf{2.9}  (0.7)&	13.4      &   -4.1    &	5.8  & N? & \\
                  & 2 & -121.4 & \bf{6.6}  (0.2)&	9.7       &   -9.2    &	2.9  & N? & SA/off-beam? \\
                  & 3 & -117.0 & 6.4            &\bf{9.8} (0.2)&      -13.7   &	3.0  & F  & \\
                  & 4 & -79.1  & 5.3       (0.3)&	11.0 (0.3)&   -       &	3.8  & ?  & ambiguous \\
G343.5024-00.0145 & 1 & -27.7  & \bf{2.8}  (0.7)&	13.5      &   -0.7    &	5.9  & N  & \\
                  & 2 & -4.1   & \bf{0.5}  (1.4)&	15.8      &   -0.1    &	8.0  & N  & \\
G343.5213-00.5171 & 1 & -34.9  & 3.3       (0.6)&	13.0 (0.6)&   -       &	5.5  & ?  & ambiguous \\
                  & 2 & -19.6  & 2.1       (0.9)&	14.2 (0.9)&   -       &	6.5  & ?  & ambiguous \\
                  & 3 & -78.6  & 5.3            &\bf{11.0}(0.3)&      -99.3   &	3.8  & F  & \\
G343.5526+00.3705 & 1 & -119.0 & \bf{6.5}  (0.2)&	9.8       &   42.0    &	2.9  & N? & strongest? \\
                  & 2 & -8.7   & 1.1            &\bf{15.2}(1.2)&      98.3    &	7.5  & F  & \\
G343.5928-00.1523 & 1 & -27.3  & \bf{2.8}  (0.8)&	13.5      &   -7.4    &	5.9  & N  & SA/double? \\
                  & 2 & -32.7  & 3.1            &\bf{13.2}(0.7)&      -35.1   &	5.6  & F  & \\
                  & 3 & -116.6 & 6.4            &\bf{9.9} (0.2)&      -26.3   &	3.0  & F  & \\
                  & 4 & -15.8  & 1.8            &\bf{14.5}(1.0)&      -38.5   &	6.8  & F? & \\
                  & 5 & -3.6   & \bf{0.5}  (1.4)&	15.8      &   -1.3    &	8.0  & N  & \\
                  & 6 & -99.1  & 5.9            &\bf{10.4}(0.3)&      -27.6   &	3.3  & F  & \\
G343.6489-00.1842 & 1 & -27.4  & \bf{2.8}  (0.8)&	13.5      &   -9.0    &	5.9  & N  & SA/double? \\
                  & 2 & -31.9  & \bf{3.1}  (0.7)&	13.2      &   -10.0   &	5.6  & N  & \\
                  & 3 & -15.9  & 1.8            &\bf{14.5}(1.0)&      -46.6   &	6.8  & F  & \\
G343.7225+00.8327 &   & -18.9  & \bf{2.1}  (0.9)&	14.2      &   30.5    &	6.5  & N? & \\
G344.7291-00.5689 & 1 & -25.7  & \bf{2.8}  (0.8)&	13.6      &   -27.8   &	5.9  & N  & strongest? \\
                  & 2 & -17.6  & \bf{2.1}  (1.0)&	14.3      &   -20.9   &	6.5  & N  & \\
                  & 3 & -32.9  & \bf{3.3}  (0.7)&	13.1      &   -32.8   &	5.4  & N  & \\
G344.8889+01.4349 &   & -13.4  & \bf{1.7}  (1.1)&	14.7      &   42.6    &	6.9  & N? & \\
G345.3307+01.4361 &   & -14.2  & \bf{1.8}  (1.1)&	14.6      &   45.1    &	6.8  & N? & \\
G345.3766-00.9023 &   & -21.2  & \bf{2.5}  (0.9)&	14.0      &   -39.4   &	6.2  & N  & \\
G345.5043+00.3480 & 1 & -17.8  & \bf{2.2}  (1.0)&	14.3      &   13.4    &	6.4  & N? & \\
                  & 2 & -38.1  & \bf{3.7}  (0.6)&	12.8      &   22.5    &	5.0  & N? & \\
G345.9561+00.6123 & 1 & -11.4  & \bf{1.5}  (1.3)&	14.9      &   16.0    &	7.0  & N  & strongest? \\
                  & 2 & -21.9  & \bf{2.6}  (0.9)&	13.9      &   27.8    &	6.0  & N  & \\
G346.2331-00.3197 &   & -11.5  & 1.6            &\bf{14.9}(1.3)&      -83.1   &	7.0  & F? & \\
G347.6327+00.2551 & 1 & -85.7  & 5.9            &\bf{10.7}(0.3)&      47.6    &	3.0  & F  & structure \\
                  & 2 & -73.4  & 5.5            &\bf{11.1}(0.3)&      49.4    &	3.3  & F  & \\
G347.8349+00.0849 & 1 & -66.7  & 5.3            &\bf{11.3}(0.3)&      16.7    &	3.5  & F  & \\
                  & 2 & -8.7   & 1.4            &\bf{15.2}(1.6)&      22.5    &	7.2  & F  & structure \\
G347.8646+00.0189 &   & -30.9  & 3.5            &\bf{13.1}(0.8)&      4.3     &	5.1  & F  & \\
G347.8903-00.2796 & 1 & -93.3  & 6.1            &\bf{10.5}(0.2)&      -51.2   &	2.8  & F  & \\
                  & 2 & -21.0  & \bf{2.7}  (1.0)&	13.9      &   -13.2   &	5.9  & N? & \\
G348.0489+00.2322 & 1 & -9.0   & \bf{1.4}  (1.6)&	15.2      &   5.7     &	7.1  & N  & SA/double? \\
                  & 2 & -12.1  & \bf{1.8}  (1.4)&	14.8      &   7.3     &	6.7  & N  & \\
                  & 3 & -0.9   & 0.2            &\bf{16.5}(2.3)&      66.9    &	8.3  & F? & \\
                  & 4 & -88.2  & 6.0            &\bf{10.6}(0.2)&      43.0    &	2.9  & F  & strongest? \\
G348.0679+00.2043 & 1 & -90.5  & 6.1            &\bf{10.6}(0.2)&      37.8    &	2.8  & F  & strongest? \\
                  & 2 & -8.7   & \bf{1.4}  (1.6)&	15.2      &   5.0     &	7.1  & N? & SA/off-beam? \\
                  & 3 & -12.3  & \bf{1.9}  (1.4)&	14.8      &   6.8     &	6.7  & N? & \\
G348.2943+00.6365 & 1 & -8.0   & \bf{1.3}  (1.7)&	15.3      &   14.4    &	7.2  & N  & \\
                  & 2 & -28.8  & 3.5            &\bf{13.2}(0.8)&      146.6   &	5.2  & F  & \\
G348.2949-00.5160 &   & -19.5  & 2.7            &\bf{14.0}(1.1)&      -126.1  &	5.9  & F  & \\
G348.5107-01.1199 & 1 & -14.4  & 2.2            &\bf{14.5}(1.3)&      -283.5  &	6.4  & F  & \\
                  & 2 & -32.9  & 3.8            &\bf{12.9}(0.7)&      -252.2  &	4.9  & F  & \\
G348.5547-00.9856 & 1 & -14.7  & 2.2            &\bf{14.5}(1.3)&      -249.5  &	6.4  & F  & \\
                  & 2 & -6.1   & \bf{1.1}  (1.9)&	15.6      &   -18.9   &	7.5  & N  & \\
                  & 3 & -29.2  & 3.5            &\bf{13.1}(0.8)&      -225.4  &	5.1  & F  & SA/weak \\
                  & 4 & -33.2  & 3.8            &\bf{12.9}(0.7)&      -221.9  &	4.8  & F  & \\
G348.5625-00.9328 & 1 & -13.7  & 2.1            &\bf{14.6}(1.4)&      -237.7  &	6.5  & F  & SA \\
                  & 2 & -33.6  & 3.8            &\bf{12.8}(0.7)&      -208.4  &	4.8  & F  & weak \\
G348.7121+00.3279 & 1 & -6.6   & \bf{1.1}  (1.8)&	15.5      &   6.3     &	7.4  & N  & \\
                  & 2 & -105.5 & \bf{6.5}  (0.2)&	10.2      &   37.2    &	2.5  & N? & \\
                  & 3 & -65.3  & 5.4            &\bf{11.3}(0.3)&      64.7    &	3.4  & F  & \\
G348.7250-01.0435 &   & -12.7  & \bf{2.0}  (1.4)&	14.7      &   -36.4   &	6.6  & N  & \\
G348.8683-00.5534 &   & -28.0  & 3.5            &\bf{13.2}(0.8)&      -127.5  &	5.1  & F  & \\
G348.8922-00.1787 & 1 & -98.4  & 6.4            &\bf{10.3}(0.2)&      -32.1   &	2.6  & F  & \\
                  & 2 & -118.3 & 6.8            &\bf{ 9.9}(0.2)&      -30.9   & 2.2  & F  & \\
G349.0944+00.0807 & 1 & -23.5  & \bf{3.2}  (1.0)&	13.5      &   4.5     &	5.4  & N  & \\
                  & 2 & -77.3  & 5.8            &\bf{10.8}(0.3)&      15.2    &	3.0  & F  & strongest? \\
G349.1772+00.0277 & 1 & -77.9  & 5.9            &\bf{10.8}(0.3)&      5.2     &	2.9  & F  & strongest? \\
                  & 2 & -97.9  & 6.4            &\bf{10.3}(0.2)&      5.0     &	2.5  & F  & \\
G349.2169+00.1450 & 1 & -64.1  & 5.4            &\bf{11.3}(0.3)&      28.6    &	3.3  & F  & \\
                  & 2 & -81.2  & 6.0            &\bf{10.7}(0.3)&      27.1    &	2.9  & F  & \\
                  & 3 & -41.1  & 4.4            &\bf{12.3}(0.6)&      31.1    &	4.3  & F  & \\
G349.3514-00.2114 & 1 & -26.7  & 6.1            &\bf{10.6}(0.9)&      -39.1   &	2.8  & F  & strongest? \\
                  & 2 & -13.8  & 2.2            &\bf{14.5}(1.4)&      -53.5   &	6.4  & F  & off-beam \\
                  & 3 & -83.6  & 3.5            &\bf{13.2}(0.2)&      -48.7   &	5.1  & F  & \\
G349.7948+00.9357 &   & -4.3   & 0.9            &\bf{15.9}(2.3)&      259.7   &	7.7  & F  & \\
G316.8119-00.0578\footnotemark[1] \footnotetext{$^1$N$_{2}$H$^{+}$ maser line being used instead of $^{13}$CO line.} & & -39.5 & \bf{2.8} (0.6)& 9.6  & -2.8  & 6.7 & N? & \\
G318.9476-00.1963\footnotemark[1] & & -36.1 & 2.5      & \bf{10.3}(0.6) & -35.3 & 6.8 &  F? & \\
G323.7408-00.2631\footnotemark[1] & & -51.5 & \bf{3.4} (0.5)& 10.3      & -15.6 & 6.1 & N  & \\
G331.2788-00.1887\footnotemark[1] & & -90.5 & \bf{5.4} (0.4)& 9.5       & -17.8 & 4.6 & N  & \\
G332.7250-00.6201\footnotemark[1] & & -50.6 & \bf{3.4} (0.5)& 11.7      & -36.8 & 5.7 & N? & \\
\end{mpsupertabular}
}
\end{center}
\small {The first column shows the \textit{MSX} object name in
Galactic co-ordinates with any weaker line-of-sight components labeled
1-n. All co-ordinates are taken from the second release of the
\textit{MSX} point source catalogue. The second column gives the
v$_{LSR}$ of the $^{13}$CO component(s), or central N$_{2}$H$^{+}$
masing line. Columns three and four are the calculated near and far
distances using the standard Galactic rotation parameters ($R_{0} =
8.5~$kpc, $\Theta_{0} = 220~$km ~s$^{-1}$) of \citet{9}. If the
kinematic distance ambiguity can be solved, the correct distance is
labelled with bold font along with a typical $\pm$10 km s$^{-1}$
peculiar velocity distance error. Columns five and six are the scale
height and distance to the Galactic centre (R$_{gc}$) of the source or
component respectively. The final column shows the adopted distance
when compared to the \hi\ data, where N and F represents near and
far-distances respectively, with N? and F? representing the most
probable solution to slightly questionable spectra. A N? generally
represents a dip in \hi\ is present, but much less than 15-20\, or is
complicated via velocity structure in $^{13}$CO. A F? generally
represents a dip may be present in \hi\, but could be atributed to
noise, or is shifted from the $^{13}$CO peak position. A lone question
mark denotes a still unsolved distance ambiguity. Any line profile
problems are briefly noted in the final column.  Notes index:
SA/Off-beam/double sources have potential absorption at the chosen
velocity component due to self absorption, off-beam objects, or could
be considered to have double-peaked profiles. Structure indicates the
line profile is not a typical gaussian. Strongest? represents a
questionable strongest/best component line within the spectra which
may or may not be the true potential MYSO from the \textit{MSX}
images. Ambiguous sources are those which still cannot be solved with
this method}

\newpage

\vspace{1.0cm}
\begin{center}
\tablefirsthead{
   \multicolumn{9}{c}{Table A2. $^{13}$CO 1-0 Kinematic parameters of \textit{MSX} sources geometrically solved with no distance ambiguity}\\
  \hline
     \hline
     \multicolumn{1}{c}{Source Name} &  & \multicolumn{1}{c}{V$_{LSR}$} & \multicolumn{1}{c}{D$_{near}$} & \multicolumn{1}{c}{D$_{far}$} & \multicolumn{1}{c}{z} & \multicolumn{1}{c}{R$_{gc}$} & Adopted & Notes \\ 
     & & (km s$^{-1}$) & (kpc) & (kpc) & (pc) & (kpc) & Distance & \\
     \hline}

\tablelasttail{\hline}
\begin{mpsupertabular}{llccccccl}
G281.5961-00.1362 & & 20.8 & - &   6.1(1.0) & -14.5  & 9.4  & - & n/a \\
G282.2992-00.7763 & & 5.5  & - &   4.5(1.2) & -61.0  &  8.7 & - & n/a \\
G290.4756-00.3681 & & 18.8 & - &   7.9(0.9) & -50.8  &  9.4 & - & n/a \\
G295.4820-00.4464 & & 41.6 & - &  11.2(1.0) & -87.3  & 10.8 & - & n/a \\
G296.4038-01.0192 & & 27.6 & - &  10.1(0.9) & -179.7 &  9.9 & - & n/a \\
G303.9973+00.2800 & & 30.7 & - &  12.2(1.0) & 59.6   & 10.2 & - & n/a \\
G306.0951-00.4778 & & 41.7 & - &  13.7(1.1) & -114.2 & 11.1 & - & n/a \\
G306.1260-01.1371 & & 1.5  & - &  10.1(0.8) & -200.5 &  8.6 & - & n/a \\
G315.4195-00.3140 & & 4.5  & - &  12.5(0.8) & -68.5  &  8.8 & - & n/a \\
G325.5964+00.0912 & & 32.6 & - &  17.5(1.5) & 27.9   & 11.5 & - & n/a \\
\end{mpsupertabular}
\end{center}

\small {The first column shows the \textit{MSX} object name in
Galactic co-ordinates. All co-ordinates are taken from the second
release of the \textit{MSX} point source catalogue. The second column
gives the v$_{LSR}$ of the $^{13}$CO component. Columns three and four
are the calculated near and far distances using the standard Galactic
rotation parameters ($R_{0} = 8.5~$kpc, $\Theta_{0} = 220~$km
~s$^{-1}$) of \citet{9} with a typical $\pm$10 km s$^{-1}$ peculiar
velocity distance error. Columns five and six are the scale height and
distance to the Galactic centre (R$_{gc}$) of the source. As these
objects are geometrically solved, there are no near distance solutions
given.}

\twocolumn

\label{lastpage}

\end{document}